\documentclass[12pt, draftclsnofoot, onecolumn]{IEEEtran}
\usepackage{stfloats}
\usepackage{float}
\usepackage{amsthm}
\usepackage{graphicx}
\usepackage{enumerate}
\usepackage{amsmath}
\usepackage{amssymb}
\usepackage{amsfonts}
\usepackage{filecontents}
\usepackage{xcolor}
\usepackage{bm}
\usepackage{cite}
\usepackage[colorlinks,linkcolor=blue,citecolor=blue]{hyperref}
\theoremstyle{definition}
\newtheorem{problem}{Problem}
\newtheorem{definition}{Definition}
\newtheorem{lemma}{Lemma}
\newtheorem{theorem}{Theorem}
\newtheorem{remark}{Remark}
\newtheorem{corollary}{Corollary}

\ifCLASSOPTIONcompsoc
\usepackage[caption=false,font=normalsize,labelfon
t=sf,textfont=sf,subrefformat=parens,labelformat=parens]{subfig}
\else
\usepackage[caption=false,font=footnotesize,subrefformat=parens,labelformat=parens]{subfig}
\fi

\title{Computation Over NOMA: Improved Achievable Rate Through Sub-Function Superposition}
\author{Fangzhou Wu, Li Chen, Nan Zhao, \IEEEmembership{Senior Member, IEEE,} Yunfei Chen, \IEEEmembership{Senior Member, IEEE,}\\ F. Richard Yu, \IEEEmembership{Fellow, IEEE,} and Guo Wei
	\thanks{F. Wu, L. Chen and G. Wei are with Department of Electronic Engineering and Information Science, University of Science and Technology of China, Hefei, Anhui 230027. (e-mail: fangzhouwu@outlook.com, \{chenli87, wei\}@ustc.edu.cn).}
	\thanks{N. Zhao is with the School of Info. and Commun. Eng., Dalian University of Technology, Dalian 116024, China, and also with National Mobile Communications Research Laboratory, Southeast University, Nanjing 210096, China. (e-mail:zhaonan@dlut.edu.cn).}
	\thanks{Y. Chen is with the School of Engineering, University of Warwick, Coventry CV4 7AL, U.K. (e-mail: Yunfei.Chen@warwick.ac.uk).}
	\thanks{F.R. Yu is with the Department of Systems and Computer Engineering, Carleton University, Ottawa, ON, K1S 5B6, Canada. (email: richard.yu@carleton.ca).}
}

\begin{document}
	\maketitle
	
	\begin{abstract}
		Massive numbers of nodes will be connected in future wireless networks. This brings great difficulty to collect a large amount of data. Instead of collecting the data individually, computation over multi-access channel (CoMAC) provides an intelligent solution by computing a desired function over the air based on the signal-superposition property of wireless channel. To improve the spectrum efficiency in conventional CoMAC, we propose the use of non-orthogonal multiple access (NOMA) for functions in CoMAC. The desired functions are decomposed into several sub-functions, and multiple sub-functions are selected to be superposed over each resource block (RB). The corresponding achievable rate is derived based on sub-function superposition, which prevents a vanishing computation rate for large numbers of nodes. In order to gain more insights, we further study the limiting case when the number of nodes goes to infinity. An exact expression of the rate is derived which provides a lower bound on the computation rate. Compared with existing CoMAC, the NOMA-based CoMAC (NOMA-CoMAC) not only achieves a higher computation rate, but also provides an improved non-vanishing rate. Furthermore, the diversity order of the computation rate of NOMA-CoMAC is derived, and it shows that the system performance is dominated by the node with the worst channel gain among these sub-functions in each RB.
	\end{abstract}
	
	
	\section{Introduction}
	In 5G and Internet of Things, tens of billions of devices are expected to be deployed to serve our societies\cite{al2015internet,gupta2015survey,vaezi2018multiple}. With such an enormous number of nodes, the collection of data using the conventional multi-access schemes is impractical because this would result in excessive network latency with limited radio resources.

	To solve the problem, a promising solution is computation over multi-access channel (CoMAC). It exploits the signal-superposition property of wireless channel by computing the desired function through concurrent node transmissions\cite{goldenbaum2015nomographic,goldenbaum2013robust,abari2016over,goldenbaum2014channel,kortke2014analog,zhu2018over,wu2019experimental,nazer2007computation,appuswamy2014computing,erez2005lattices,wu2018experimental,nazer2011compute,jeon2014computation,Jeon2016Opportunistic,wang2015interactive,goldenbaum2014computation}. Many networks computing a class of nomographic functions of the distributed data can employ CoMAC\cite{goldenbaum2015nomographic}. For example, wireless sensor networks can use the framework of CoMAC since it only aims to obtain a functional value of the sensor readings (e.g., arithmetic mean, polynomial or the number of active nodes) instead of requiring all readings from all sensors. 
	
	Analog CoMAC was first studied in \cite{goldenbaum2013robust,abari2016over,goldenbaum2014channel,kortke2014analog,zhu2018over,wu2019experimental}, where pre-processing at each node and post-processing at the fusion center were used to fight fading and compute functions. The designs of pre-processing and post-processing used to compute linear and non-linear functions have been proposed in\cite{abari2016over}, and the effect of channel estimation error was characterized in \cite{goldenbaum2014channel}. In order to verify the feasibility of analog CoMAC in practice, software defined radio was built in \cite{kortke2014analog}. A multi-function computation method has been presented in \cite{zhu2018over}, which utilized a multi-antenna fusion center to collect data transmitted by a cluster of multi-antenna multi-modal sensors. Also, the authors in \cite{wu2019experimental} studied how to compute multiple functions over-the-air with antennas arrays at devices and the access point, where different linear
	combinations with arbitrary coefficients for the Gaussian sources were
	computed. In summary, the simple analog CoMAC has led to an active area focusing on the design and implementation techniques for receiving a desired function.

	Since analog CoMAC is not robust to noise, digital CoMAC was proposed to use joint source-channel coding in \cite{nazer2007computation,appuswamy2014computing,erez2005lattices,nazer2011compute,jeon2014computation,Jeon2016Opportunistic,wang2015interactive,goldenbaum2014computation,wu2018experimental,wu2018computation} to improve equivalent signal-to-noise ratio (SNR). The potential of linear source coding was discussed in \cite{nazer2007computation}, and its application was presented in \cite{appuswamy2014computing} for CoMAC. Compared with linear source coding, nested lattice coding can approach the performance of a standard random coding\cite{erez2005lattices}. The lattice-based CoMAC was extended to a general framework in \cite{nazer2011compute} for relay networks with linear channels and additive white Gaussian noise (AWGN). In order to combat non-uniform fading, a uniform-forcing transceiver design was given in \cite{wu2018experimental}. Achievable computation rates were given in \cite{nazer2011compute,jeon2014computation,Jeon2016Opportunistic} for digital CoMAC through theoretical analysis.

	One serious issue in digital CoMAC is the vanishing rate as the number of nodes increases when the fading MAC is considered. In order to prevent the vanishing rate, a narrow-band CoMAC (NB-CoMAC) with opportunistic computation has been studied in \cite{Jeon2016Opportunistic}. A wide-band CoMAC (WB-CoMAC) has been extended in \cite{wu2018computation} against both frequency selective fading and the vanishing computation rate.
	
	To the best of our knowledge, all the aforementioned CoMAC works only considered the use of orthogonal multiple access (OMA) for functions by transmitting the function in different resource blocks (RBs), i.e, time slots or sub-carriers. Non-orthogonal multiple access (NOMA) is well-known for improving spectrum efficiency but has never been considered in CoMAC\cite{ding2017survey,dai2018survey,wang2006comparison}. For example, the authors in
	\cite{wang2006comparison} compared NOMA and OMA in the uplink, which showed that NOMA achieves higher ergodic sum rates while the fairness of nodes has been considered. NOMA based on user pairing was considered in \cite{ding2016impact,yang2016general,sedaghat2018user} to ease successive interference cancellation (SIC) caused by the superposition transmission of too many nodes at the base station. Different from NOMA for information transmission, NOMA-based CoMAC (NOMA-CoMAC) superposes multiple functions instead of different bit sequences from different nodes in each RB. Also, nodes with
	disparate channel conditions are allowed to be served simultaneously in conventional NOMA to improve the performance, whereas the node with poor channel condition only makes the computation rate vanishing in NOMA-CoMAC system since the function computed by nodes requires the uniform fading. It suggests that the direct use of NOMA in CoMAC system is not suitable.

	Motivated by the above observations, in this work, we propose a NOMA-CoMAC system through the division, superposition, SIC and reconstruction of the desired functions. Analytical expression for the achievable computation rate with sub-function superposition is derived based on nested lattice coding\cite{jeon2014computation,Jeon2016Opportunistic,wang2015interactive,goldenbaum2014computation}. Furthermore, several limiting cases are considered to characterize the lower bound of the computation rate and the diversity order. Our contributions can be summarized as follows: 
	
	\begin{itemize}
		\item \emph{Novel NOMA-CoMAC}. We propose a NOMA-CoMAC system with sub-function superposition. Unlike NOMA systems for information transmission, NOMA-CoMAC decomposes the desired functions into several sub-functions, superposes these sub-functions with large equivalent channel gains in each RB, executes the process of SIC and reconstructs the desired functions at the fusion center.
		\item \emph{Improved computation rate}. The analytical expression of the computation rate of NOMA-CoMAC is derived. Using the closed-form expression, the power allocated to each node can be calculated directly with low computational complexity. Compared with the conventional CoMAC schemes, both achievable computation rate and non-vanishing rate with massive nodes are improved.
		\item \emph{Limiting cases}. We characterize the lower bound of the computation rate with an exact expression as the number of nodes goes to infinity. It provides a straightforward way to evaluate the system performance. As the power of each node goes to infinity, we obtain the diversity order of the computation rate of NOMA-CoMAC. It shows that the node with the worst channel gain among these sub-functions in each RB plays a dominant role.
	\end{itemize}
	
	The rest of the paper is organized as follows. Section \ref{Preliminaries and Problem Statement} introduces the definitions and the existing results of NB-CoMAC and WB-CoMAC. The system model of NOMA-CoMAC is further given. In Section \ref{Main Results}, we summarize the main results of this paper and compare them with the conventional CoMAC schemes. Section \ref{Design of NOMA-CoMAC} presents the proposed NOMA-CoMAC with sub-function superposition in detail and analyzes the computation rate. Section \ref{Power Control for NOMA-CoMAC} focuses on the performance of the proposed NOMA-CoMAC, which includes the power control and outage. Simulation results and the corresponding discussion are presented in Section \ref{Simulation Results and Discussions}, and conclusions are given in Section \ref{Conclusion}.

	\section{System Model}
	\label{Preliminaries and Problem Statement}
	
	We introduce two typical CoMAC frameworks in this section, and review the main results of several previous works. We define $ \mathsf{C}^+(x)=\max\left\lbrace \frac{1}{2}\log(x),0\right\rbrace$ and $ \lceil  x \rceil=\min\left\lbrace n\in\mathbb{Z}|x\le n \right\rbrace  $ as the ceiling function. Let $  [1:n] $ denote a set $ \left\lbrace1,2,\cdots,n\right\rbrace$ and $ (\cdot)^{\mathrm{T}} $ represent the transpose of a vector or matrix. For a set $ \mathcal{A} $, $ \left|\mathcal{A}\right|  $  denotes the cardinality of $ \mathcal{A} $. Let the entropy of a random variable $ A $ be $ H(A) $ and $ \mathrm{diag}\left\lbrace a_1, a_2, \cdots, a_n\right\rbrace  $ denote the diagonal matrix of which the diagonal elements are from $ a_1 $ to $ a_n $. A set $ \left\lbrace x_1, x_2, \cdots, x_N \right\rbrace  $ is written as $ \left\lbrace x_i \right\rbrace_{i\in[1:N]}  $ or $ \left\lbrace x_i \right\rbrace_{i=1}^{N} $ for short.
	
	\subsection{Narrow-Band CoMAC}
	\begin{figure}
		\centering
		\includegraphics[width=0.7\linewidth]{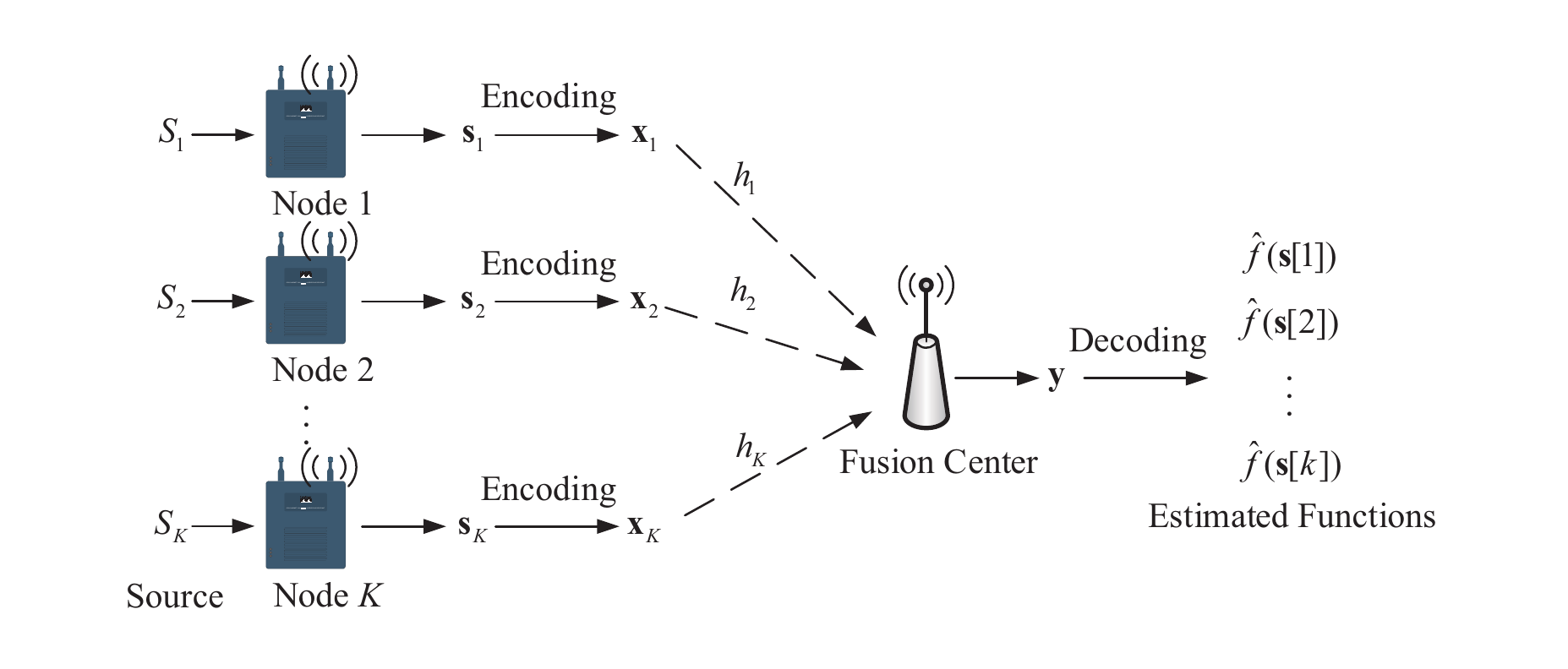}
		\caption{Framework of narrow-band CoMAC}
		\label{fig:system}
	\end{figure}
	The framework of NB-CoMAC is shown as Fig.~\ref{fig:system}, where each node draws data from a corresponding random source for $ T_d $ times and obtains a length-$ T_d $ data vector. After encoding the data vector, the fusion center computes a desired function as all nodes transmit theirs data vector simultaneously. 
	
	First of all, we define the data matrix to describe the data from $ K $ nodes during $ T_d $ time slots.

	\begin{definition}[Data Matrix]\label{Sources}\label{Discrete sample vector}
		A data matrix $ \bm{\mathrm{S}}\in\mathbb{C}^{T_d\times K}$ represents the data from $ K $ nodes during $ T_d $ time slots. It is expressed as
		\begin{align}
			\bm{\mathrm{S}}&=\begin{bmatrix}
				s_{1}[1] & \cdots & s_{1}[j] & \cdots & s_{1}[T_d] \\ 
				\vdots &  & \vdots &  & \vdots \\ 
				s_{i}[1] & \cdots & s_{i}[j] & \cdots & s_{i}[T_d] \\ 
				\vdots &  & \vdots &  & \vdots \\ 
				s_{K}[1] & \cdots & s_{K}[j] & \cdots & s_{K}[T_d]
			\end{bmatrix} \nonumber\\
			&=\begin{bmatrix}
				{\bm{\mathrm{s}}[1]}^{\mathrm{T}} & \cdots & {\bm{\mathrm{s}}[j]}^{\mathrm{T}} & \cdots & {\bm{\mathrm{s}}[T_d]}^{\mathrm{T}}
			\end{bmatrix} \nonumber\\
			&=\begin{bmatrix}
				\bm{\mathrm{s}}_1^{\mathrm{T}} & \cdots & \bm{\mathrm{s}}_i^{\mathrm{T}} & \cdots & \bm{\mathrm{s}}_K^{\mathrm{T}}
			\end{bmatrix}^{\mathrm{T}},
		\end{align}
		where $ j\in[1:T_d] $, $ i\in[1:K] $, $ s_{i}[j] $ is the $ j $-th data of the $ i $-th node from the random source $ S_i $, $ \bm{\mathrm{s}}[j]=[s_1[j],\cdots,s_K[j]] $ is the $ j $-th data of all $ K $ nodes and $ \bm{\mathrm{s}}_i=\left[s_i[1],\cdots,s_i[T_d] \right] $ is the data vector of node $ i $. Each data $ s_i[j]$ belongs to $[0:p-1] $, which means it is mapped to a number between 0 and $ p-1 $ through quantization. Let $ \bm{\mathrm{s_r}}=\left[ S_1, S_2, \cdots, S_K\right]  $ be a random vector associated with a joint probability mass function $ p_{\bm{\mathrm{s_r}}}(\cdot) $ as $ \bm{\mathrm{s}}[j]$ is independently drawn from $ p_{\bm{\mathrm{s_r}}}(\cdot) $.
	\end{definition}

	A function $ f(\bm{\mathrm{s_r}}) $ with respect to the random source vector $ \bm{\mathrm{s_r}} $ is called the desired function. Its definition is given as follows.
	\begin{definition}[Desired Function]\label{Desired Function D}
		For all $ j\in[1:T_d] $, every function
		\begin{equation}\label{Desired Function}
			f(s_1[j],s_2[j],\cdots,s_K[j])=f(\bm{\mathrm{s}}[j])
		\end{equation}
		that is computed at the fusion center is called a desired function where $\bm{\mathrm{s}}[j]$ is independently drawn from $ p_{\bm{\mathrm{s_r}}}(\cdot) $ (See Definition \ref{Sources}). Every function $ f(\bm{\mathrm{s}}[j]) $ can be seen as a realization of $  f(\bm{\mathrm{s_r}})  $. Thus, it has $ T_d $ functions when each node gets data from each random source for $ T_d $ times. 
	\end{definition}
	
	In order to be robust against noise, we use sequences of nested lattice codes\cite{nazer2011compute} throughout this paper. Based on this coding, the definitions of encoding and decoding are given as follows.
	
	\begin{definition}[Encoding \& Decoding]\label{Encoding & Decoding}
		Let $ \bm{\mathrm{s}}_i $ denote the data vector for the $ i $-th node whose length is $ T_d $ (see Definition \ref{Discrete sample vector}). Denote $ \bm{\mathrm{x}}_i=[x_i[1],x_i[2],\cdots,x_i[n]] $ as the length-$ n $ transmitted vector for node $ i $. The received vector with length $ n $ is given by $ \bm{\mathrm{y}}=[y[1],y[2],\cdots,y[n]] $ at the fusion center. Assuming a block code with length $ n $, the encoding and decoding functions can be expressed as follows.
		\begin{itemize}
			\item \emph{Encoding Functions}: the univariate function $ \bm{\mathcal{E}}_i(\cdot) $ which generates $ \bm{\mathrm{x}}_i=\bm{\mathcal{E}}_i(\bm{\mathrm{s}}_i) $ is an encoding function of node $ i $. It maps $ \bm{\mathrm{s}}_i $ with length $ T_d $ to a transmitted vector $ \bm{\mathrm{x}}_i $ with length $ n $ for node $ i $.
			\item \emph{Decoding Functions}: the decoding function $ \bm{\mathcal{D}}_j(\cdot) $ is used to estimate the $ j $-th desired function $  f(\bm{\mathrm{s}}[j])  $, which satisfies $ \hat{f}(\bm{\mathrm{s}}[j])=\bm{\mathcal{D}}_j(\bm{\mathrm{y}}) $. It implies the fusion center obtains $ T_d $ desired functions depending on the whole received vector with length $ n $.
		\end{itemize}
	\end{definition}

	Considering the block code with length $ n $, the definition of computation rate\cite{Jeon2016Opportunistic,jeon2014computation,goldenbaum2015nomographic,goldenbaum2014computation} can be given as follows.
	\begin{definition}[Computation rate]\label{Computation rate}
		The computation rate specifies how many function values can be computed per channel use within a predefined accuracy. It can be written as $ R=\lim\limits_{n\rightarrow \infty}\frac{T_d}{n}H(f(\bm{\mathrm{s_r}})) $ where $ T_d $ is the number of function values (see Definition \ref{Desired Function D}), $ n $ is the length of the block code and $ H(f(\bm{\mathrm{s_r}})) $ is the entropy of $ f(\bm{\mathrm{s_r}})  $. Apart from this, $ R $ is achievable only if there is a length-$ n $ block code so that the probability $ \Pr\left( \bigcup_{j=1}^{T_d}\left\lbrace \hat{f}(\bm{\mathrm{s}}[j]\neq f(\bm{\mathrm{s}}[j]))\right\rbrace\right)  \rightarrow0 $ as $ n $ increases.
	\end{definition}

	\subsection{Wide-Band CoMAC}\label{Wide-Band CoMAC}
	\begin{figure}
		\centering
		\includegraphics[width=0.7\linewidth]{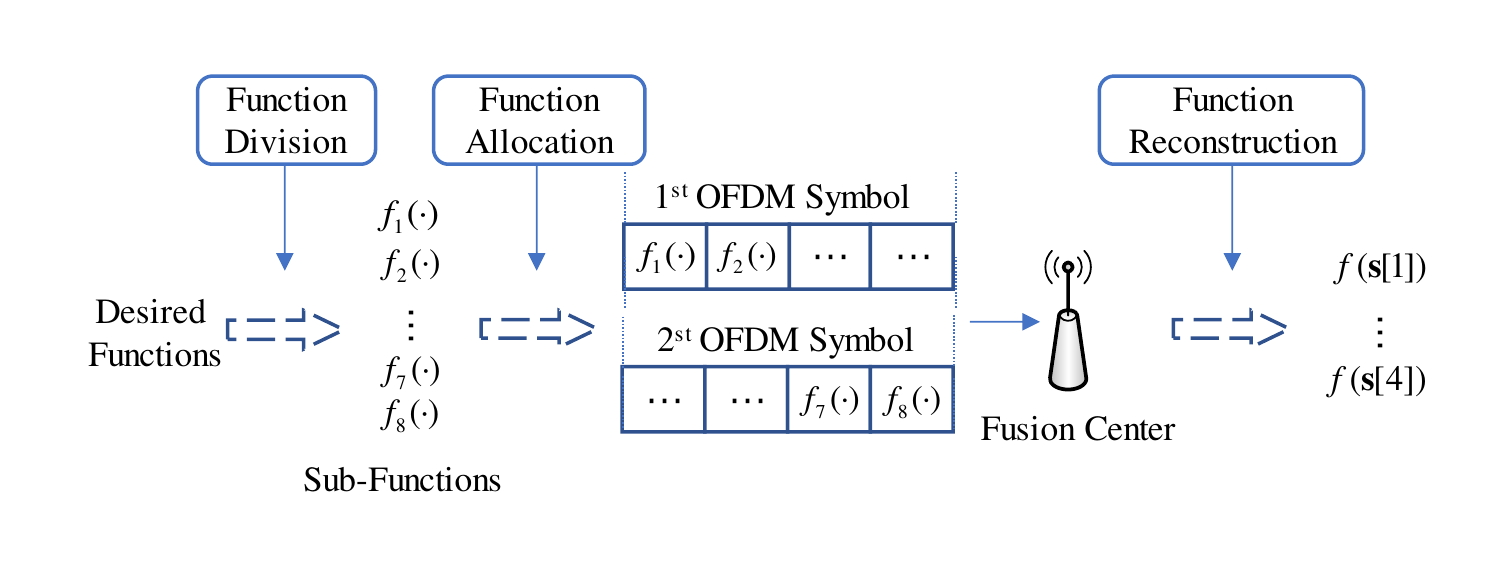}
		\caption{Framework of wide-band CoMAC}
		\label{fig:ofdm-comac}
	\end{figure}
	
	The framework of CoMAC was adopted to a wide-band MAC which aims to improve the computation rate over NB-CoMAC. The desired function as a whole cannot be allocated into several sub-carriers. Thus, it divided the desired function into sub-functions transmitted as bit sequences, and allocated sub-functions to each sub-carrier so as to provide a improved non-vanishing computation rate. The framework of WB-CoMAC is shown in Fig.~\ref{fig:ofdm-comac}.
	
	A desired function is computed through the simultaneous transmission of all nodes. If only a subset of nodes participates in the computation, the function computed by a subset of nodes is called the sub-function. The sub-function is only part of the desired function. Assuming that a sub-function is computed by $ M $ chosen nodes and the number of all nodes is $ K $, the desired function is split into $ B=\frac{K}{M} $ parts.

	\begin{definition}[Sub-Function]\label{Partitioning Function}
		Let \begin{equation}\label{key}
			\tau_u=\left\lbrace x\in[1:K]:\left|\tau_u\right|=M \right\rbrace
		\end{equation} 
		denote a set where each element $ x $ is the index from the $ M $ chosen nodes. Suppose that $ \bigcup_{u=1}^{B}\tau_u=[1:K] $ and $ \tau_u\bigcap\tau_v =\emptyset$ for all $ u,v\in[1:B] $, a function $ f\left(\left\lbrace s_i[j]\right\rbrace_{i\in\tau_u} \right) $ is said to be a sub-function if and only if there exists a function $ f_c(\cdot) $ satisfying $ f(\bm{\mathrm{s}}[j])=f_c(\lbrace f(\lbrace s_i[j]\rbrace_{i\in\tau_u})\rbrace _{u\in[1:B]})  $.
	\end{definition}

	\begin{remark}[Detachable functions]
		As studied in \cite{giridhar2005computing,wang2015interactive,goldenbaum2014computation,jeon2014computation}, CoMAC can be designed to compute different types of desired function. We focus on two typical functions, the arithmetic sum function and the type function. A function $ f(\bm{\mathrm{s}}[j])=\sum_{i=1}^{K}a_{i}s_i[j]$ is the arithmetic sum function where $ a_{i} $ is the weighting factor for node $ i $. A function $ f(\bm{\mathrm{s}}[j])=\sum_{i=1}^{K}\bm{1}_{s_i[j]=b}$ is regarded as the type function where $ \bm{1}_{(\cdot)} $ denotes the indicator function and $ b\in[0,p-1] $. Both arithmetic sum function and type function are detachable from Definition \ref{Partitioning Function}.
	\end{remark}

	\subsection{Existing Results}
	\label{Previous Works}
	The computation rates of NB-CoMAC and WB-CoMAC are given as follows.
	
	\begin{theorem}[Rate of NB-CoMAC]\label{Older}
		As shown in \cite[Theorem 1]{Jeon2016Opportunistic}, for any $ M,B\in\mathbb{N} $ satisfying $ MB=K $, the ergodic computation rate of NB-CoMAC is given by
		\begin{equation}\label{QINC}
			R=\dfrac{1}{B}\mathsf{E}\left[ \mathsf{C}^+\left( \dfrac{1}{M}+\dfrac{|h_{\mathcal{I}_{M}}|^2KP}{M\mathsf{E}\left[\dfrac{|h_{\mathcal{I}_{M}}|^2}{|h|^2} \right]}\right)\right],
		\end{equation}
		where $ K $ is the number of nodes, $ M $ is the number of the chosen nodes to compute a sub-function, $ |h_{\mathcal{I}_{M}}|^2 $  is the channel gain of the $ \mathcal{I}_{M} $-th node, $ \mathcal{I}_{M} $ is the $ M $-th element of the set of the ordered indexes of channel gains $ \left\lbrace\mathcal{I}_{i}\right\rbrace_{i\in[1:K]} $ such that $ |h_{\mathcal{I}_{1}}|^2\ge|h_{\mathcal{I}_{2}}|^2\ge\cdots\ge|h_{\mathcal{I}_{K}}|^2  $ and $ h $ is any random variable.
	\end{theorem}

	Theorem \ref{Older} considered a NB-CoMAC with flat fading, \cite{wu2018computation} expanded it to a WB-CoMAC with frequency selective fading to focus on high speed transmission.
	
	\begin{theorem}[Rate of WB-CoMAC]\label{ESubPowerINC}
		As mentioned in \cite[Corollary 2 and Eq.~(27)]{wu2018computation}, for any $ M, N\in\mathbb{N} $ satisfying $ K=BM $, the ergodic computation rate of WB-CoMAC over frequency selective fading MAC is given by
		\begin{equation}\label{key}
			R=\dfrac{1}{BN}\mathsf{E}\left[\sum_{g=1}^{N} \mathsf{C}^+\left( \dfrac{N}{M}+ \dfrac{KP|h_{\mathcal{I}_{M}^g,g}|^2}{M\mathsf{E}\left[\dfrac{|h_{\mathcal{I}_{M}^g,g}|^2}{|h|^2} \right]}\right)\right],
		\end{equation}		
		where $ N $ is the number of sub-carriers, $ |h_{\mathcal{I}_{M}^{g},g}|^2 $  is the channel gain of the $ \mathcal{I}_{M}^{g} $-th node at the $ g $-th sub-carrier and $ \mathcal{I}_{M}^{g} $ is the $ M $-th element of the set of ordered indexes of channel gains $ \left\lbrace\mathcal{I}_{i}^g\right\rbrace_{i\in[1:K]} $ at the $ g $-th sub-carrier such that $ |h_{\mathcal{I}_{1}^g,g}|^2\ge|h_{\mathcal{I}_{2}^g,g}|^2\ge\cdots\ge|h_{\mathcal{I}_{K}^g,g}|^2  $.
	\end{theorem}
	One sees that both CoMAC schemes in the above only consider the use of OMA to transmit a function in each RB. This results in low spectrum efficiency. Since NOMA can offer extra improvement in spectrum efficiency, we apply NOMA to CoMAC to improve the computation rate. Thus, we propose a NOMA-CoMAC system with sub-function superposition, where each sub-carrier can serve these sub-functions with large equivalent channel gains simultaneously. It can not only achieve a higher computation rate, but also can provide an improved non-vanishing rate with massive nodes.

	\subsection{Novel NOMA for Wide-Band MAC}
	\label{System Model and Main Results}
	The framework of WB-CoMAC discussed in Section \ref{Wide-Band CoMAC} will be used to transmit multiple functions simultaneously in each sub-carrier using NOMA. We consider an OFDM-based system with $ N $ sub-carriers during $ T_s $ OFDM symbols while the length of the block code is $ n $. In each sub-carrier, $ L $ functions are chosen to be transmitted. Then, the $ m $-th received OFDM symbol at the fusion center can be expressed as
	\begin{equation}\label{OFDM-NOMA model}
		\bm{\mathrm{Y}}[m]=\sum_{l=1}^{L}\sum_{i=1}^{K}\bm{\mathrm{V}}_i^{l}[m]\bm{\mathrm{X}}_i^{l}[m]\bm{\mathrm{H}}_i[m]+\bm{\mathrm{W}}[m],
	\end{equation}
	where $ m\in[1:T_s] $, $ T_s=\lceil  \frac{n}{N} \rceil $, $ K $ is the number of nodes, the power allocation matrix of node $ i $ is $ \bm{\mathrm{V}}_i^{l}[m]=\mathrm{diag}\left\lbrace v_{i,1}^l[m],\cdots,v_{i,N}^l[m]\right\rbrace$ whose diagonal element is the power allocated to compute the $ l $-th function at each sub-carrier, $ \bm{\mathrm{X}}_i^{l}[m]=\mathrm{diag} \left\lbrace x^l_{i,1}[m], x^l_{i,2}[m], \cdots, x^l_{i,N}[m] \right\rbrace   $ is the transmitted diagonal matrix of node $ i $ to compute the $ l $-th function, a diagonal matrix $ \bm{\mathrm{H}}_i[m]=\mathrm{diag}\left\lbrace h_{i,1}[m],\cdots h_{i,N}[m]\right\rbrace $ is the channel response matrix of which the diagonal element is the channel response of each sub-carrier for node $ i $ and the diagonal element of $ \bm{\mathrm{W}}[m] $ is identically and independently distributed (i.i.d.) complex Gaussian random noise following $ \mathcal{CN}(0,\frac{1}{N}) $. 
	
	Assuming perfect synchronization and perfect removal of inter-carrier interference, based on Eq.~\eqref{OFDM-NOMA model}, the received signal in the $ g $-th sub-carrier at the $ m $-th OFDM symbol can be given as
	\begin{equation}\label{OFDM-NOMA model of sub-carrier}
		y_g[m]=\sum_{l=1}^{L}\sum_{i=1}^{K}v_{i,g}^l[m]x_{i,g}^l[m]h_{i,g}[m]+w_{i,g}[m],
	\end{equation}
	where $ v_{i,g}^l[m] $ is the power of node $ i $ allocated in the sub-carrier $ g $ for computing the $ l $-th function, $ x_{i,g}^l[m] $ is the transmitted symbol of node $ i $ for the $ l $-th function in the $ g $-th sub-carrier from the transmitted vector $ \bm{\mathrm{x}}_i $ (See Definition \ref{Encoding & Decoding}), $ h_{i,g}[m] $ is the channel response of the sub-carrier $ g $ for node $ i $ and $ w_{i,g}[m] $ is i.i.d. complex Gaussian random noise following $ \mathcal{CN}(0,\frac{1}{N}) $.
	%
	\section{Main Results}
	\label{Main Results}
	In the OFDM-based system, all the nodes is sorted by their channel gains in each sub-carrier, and the ordered nodes are divided into $ B=\frac{K}{M}\in\mathbb{N} $ parts to compute $ B $ sub-functions. Only the first $ L \le B$ sub-functions with large equivalent channel gains are chosen to be superposed in a sub-carrier. Then, the computation rate of NOMA-CoMAC is achievable with the limit of large $ n $.
	
	\begin{theorem}[Rate of NOMA-CoMAC]\label{OMF-NOMA rate}
		For any $ M,L,N,K\in\mathbb{N} $ satisfying $ L\le B $ and $ K=BM $, the ergodic computation rate of NOMA-CoMAC over wide-band MAC is given as
		\begin{equation}\label{General Rate}
			R=\dfrac{ML}{KNT_s}\sum_{m=1}^{T_s}\left[ \sum_{g=1}^{N}\min_{i\in[1:L]}\left[ \mathsf{C}^{+}\left( \dfrac{N}{M}+\dfrac{\displaystyle N\min_{u\in\mathcal{M}_l}\left[ |h_{\mathcal{I}_u^g[m],g}[m]|^2P_{\mathcal{I}_u^g[m],g}[m]\right] }{1+N\displaystyle\sum_{j=i+1}^{L}\displaystyle\min_{u\in\mathcal{M}_j}\left[ |h_{\mathcal{I}_u^g[m],g}[m]|^2P_{\mathcal{I}_u^g[m],g}[m]\right] }\right)\right]\right],
		\end{equation}
	where $ K $ is the number of nodes, $ M $ is the number of chosen nodes to compute a sub-function, $ L $ is the number of chosen sub-functions in each sub-carrier, $ N $ is the number of sub-carriers, $ T_s=\lceil  \frac{n}{N} \rceil $ is the number of OFDM symbols, $ \mathcal{M}_x=[M(x-1)+1:Mx] $ is a set including the indexes of the chosen nodes to compute the corresponding sub-function, $ \mathcal{I}_u^g[m] $ is the $ u $-th index of the ordered indexes of $ K $ nodes in the $ g $-th sub-carrier at the-$ m $ OFDM symbol,  $ |h_{\mathcal{I}_u^g[m],g}[m]|^2 $ is the channel gain of the $ \mathcal{I}_u^g[m] $-th node in the $ g $-th sub-carrier at the $ m $-th OFDM symbol and $ P_{\mathcal{I}_u^g[m],g}[m] $ is the power allocated to node $ \mathcal{I}_u^g[m] $.
	\end{theorem}
	\begin{IEEEproof}
		Please refer to Section \ref{Opportunistic Multi-Function NOMA for OFDM-based System} for proof.
	\end{IEEEproof}
	
	\begin{remark}[Property of NOMA-CoMAC]
		Theorem \ref{OMF-NOMA rate} presents a general rate which can be used with power control. It shows that the rate of NOMA-CoMAC is determined by the sub-function with the slowest rate among the $ L $ sub-functions in every sub-carrier, since the desired function can not be reconstructed unless all sub-functions are received at the fusion center.
	\end{remark}
	
	\begin{remark}[Generalization of rates for NB-CoMAC and WB-CoMAC]
		The rate of NOMA-CoMAC in Theorem \ref{OMF-NOMA rate} generalizes the rates of NB-CoMAC and WB-CoMAC. By setting $ L=1 $ in Theorem \ref{OMF-NOMA rate}, we can obtain the rate of WB-CoMAC in Theorem \ref{ESubPowerINC}. The rate of NB-CoMAC in Theorem \ref{Older} can be also obtained by setting $ L=1,N=1 $ in Theorem \ref{OMF-NOMA rate}.
	\end{remark}
	
	%
	%
	%
	%
	%
	\begin{table*}[b]
		\renewcommand\arraystretch{1.5}
		\centering
		\caption{Summary of Rates of NB-CoMAC, WB-CoMAC and NOMA-CoMAC}
		\label{table rates}
		\resizebox{\linewidth}{!}{
			\begin{tabular}{|c|c|c|}
				\hline 
				CoMAC Scheme & Achievable Rate & Limiting Rate \\ 
				\hline 
				NB-CoMAC & $ R=\dfrac{1}{B}\mathsf{E}\left[ \mathsf{C}^+\left( \dfrac{1}{M}+\dfrac{|h_{\mathcal{I}_{M}}|^2KP}{M\varpi_{1}}\right)\right] $ \cite{Jeon2016Opportunistic} & $ R(r)=r\mathsf{C}^+\left( \dfrac{1}{rK}+\dfrac{\xi_{1-r}P}{r\varpi_1}\right) $  \\ 
				\hline 
				WB-CoMAC & $ R=\dfrac{1}{BN}\mathsf{E}\left[\displaystyle\sum_{g=1}^{N} \mathsf{C}^+\left( \dfrac{N}{M}+ \dfrac{KP|h_{\mathcal{I}_{M}^g,g}|^2}{M\varpi_{1,g}}\right)\right] $ \cite{wu2018computation} & $ R(r)=r\mathsf{C}^+\left( \dfrac{N}{rK}+\dfrac{\xi_{1-r}P}{r\varpi_1}\right) $   \\ 
				\hline 
				NOMA-CoMAC & $ R=\dfrac{2}{BN}\mathsf{E}\left[\displaystyle\sum_{g=1}^{N}\mathsf{C}^{+}\left( \dfrac{N}{M}+\dfrac{2P\frac{K}{M}|h_{\mathcal{I}_{M}^g,g}|^2 }{\Gamma+\sqrt {
						\Gamma^{2}+4
						\,{ P\frac{K}{M}}\,{ |h_{\mathcal{I}_{M}^g,g}|^2}{ \varpi_{1,g}} }} \right)\right] $ & $ R(r)=2r\mathsf{C}^{+}\left( \dfrac{N}{rK}+\dfrac{2P\xi_{1-r}\xi_{1-2r}}{r\Delta_{\rm M} +\sqrt{(r\Delta_{\rm M})^2+4r\varpi_1P\xi_{1-r}\xi_{1-2r}^2}}\right) $ \\ 
				\hline 
			\end{tabular} 
		}
	\end{table*}
	
	In the proposed scheme, we choose the first $ L $ sub-functions with large channel gains in each sub-carrier. Since the superposition transmission of too many sub-functions makes SIC at the fusion center difficult, only two sub-functions ($ L=2 $) as a pair are chosen to be transmitted in a single sub-carrier.

	\begin{corollary}[Rate of NOMA-CoMAC with average power control]\label{intra-block power alloaction}Considering an OFDM-based system where each sub-carrier serves a sub-function pair ($ L=2 $), the ergodic computation rate of NOMA-CoMAC considering the average power control, i.e, $ \mathsf{E}\left[ P_{i,g}[m]\right] \le \frac{P}{N} $ for node $ i $, can be obtained as
		\begin{equation}\label{ex-rate}
			\resizebox{!}{!}{$
				R=\dfrac{2}{BN}\mathsf{E}\left[\sum_{g=1}^{N}\mathsf{C}^{+}\left( \dfrac{N}{M}+\dfrac{2P\frac{K}{M}|h_{\mathcal{I}_{M}^g,g}|^2 }{\Gamma+\sqrt {
						\Gamma^{2}+4
						\,{ P\frac{K}{M}}\,{ |h_{\mathcal{I}_{M}^g,g}|^2}{ \varpi_{1,g}} }} \right)\right],
				$}
		\end{equation}	
		where $ \Gamma=\frac{{ |h_{\mathcal{I}_{M}^g,g}|^2}}{{ |h_{\mathcal{I}_{2M}^g,g}|^2}}{ \varpi_{2,g}}+{ \varpi_{1,g} }$ and $\varpi_{l,g}=\mathsf{E}\left[\frac{|h_{\mathcal{I}_{Ml}^g,g}|^2}{|h|^2} \right]$.
		
	\end{corollary}
	\begin{IEEEproof}
		Please refer to Section \ref{Intra-Block Power Control} and Appendix \ref{Proof of Corollary intra-block power alloaction} for proof.
	\end{IEEEproof}
	
	Corollary \ref{intra-block power alloaction} provides an easy way to allocate the power into each sub-function when average power control is considered, since the power allocated to each node can be calculated directly using the closed-form expression with low computational complexity.
	
	Similar to the previous works, the rate of NOMA-CoMAC in Corollary \ref{intra-block power alloaction} can also prevent the rate from vanishing as the number of nodes $ K $ increases. Nevertheless, the previous works only verified the non-vanishing rate through simulation and did not obtain its exact value through mathematical analysis. We characterize the lower bound of the computation rate as the limiting rate. It can be used to calculate the accurate value of the non-vanishing computation rate with given parameters.
	
	\begin{corollary}[Limiting Rate of NOMA-CoMAC]\label{Limiting Rate of OMF-NOMA}
		As $ K $ increases, the computation rate of NOMA-CoMAC approaches an exact value which is only determined by $ r=\frac{M}{K} $ and can be given as
		\begin{equation}\label{limit rate 1}
			\resizebox{!}{!}{$
				\begin{split}
				R(r)=2r\mathsf{C}^{+}\left( \dfrac{N}{rK}+\dfrac{2P\xi_{1-r}\xi_{1-2r}}{r\Delta_{\rm M} +\sqrt{(r\Delta_{\rm M})^2+4r\varpi_1P\xi_{1-r}\xi_{1-2r}^2}}\right),
				\end{split}
				$}
		\end{equation}
		where $ \Delta_{\rm M}=\varpi_1\xi_{1-2r}+\varpi_2\xi_{1-r} $, $\varpi_{l}=\mathsf{E}\left[\frac{|h_{\mathcal{I}_{Ml}}|^2}{|h|^2} \right]$, $ F_{|h|^2}(\xi_{x})= x$ and $ F_{|h|^2} $ is the cumulative distribution function (CDF) of $ |h|^2 $. For i.i.d. Rayleigh fading,  $ F_{|h|^2} $ is the CDF of the exponential distribution with parameter one, i.e.,  $ F_{|h|^2}=1-\exp(-x) $ and $ \xi_{x}=-\ln(1-x) $.
	\end{corollary}
	\begin{IEEEproof}
		Please refer to Appendix \ref{Proof of Corollary Limiting Rate of OMF-NOMA} for proof.
	\end{IEEEproof}
	
	Note that previous works only proved that the computation rate was non-vanishing through simulation, Corollary \ref{Limiting Rate of OMF-NOMA} provides the lower bound of the computation rate of NOMA-CoMAC, which is easier to evaluate the performance. Using a similar proof of Corollary \ref{Limiting Rate of OMF-NOMA}, we can obtain the limiting rates of WB-CoMAC and NB-CoMAC.
	
	\begin{remark}[Limiting Rates for NB-CoMAC and WB-CoMAC]
		No exact lower bound of the computation rates and their limiting rates are available in previous works. Hence, we derive the exact expression of these limiting rates, which can calculate the exact values of these non-vanishing rates with given parameters. Following a similar proof, the limiting rate of WB-CoMAC in Theorem \ref{ESubPowerINC} can be obtained easily as
		\begin{equation}\label{limit rate 2}
			R(r)=r\mathsf{C}^+\left( \dfrac{N}{rK}+\dfrac{\xi_{1-r}P}{r\varpi_1}\right).
		\end{equation}
		It also generalizes the limiting rate of NB-CoMAC in Theorem \ref{Older} as $ N=1 $. Unlike conventional works with respect to 
		a series of random variables and $ M $, these limiting rates are only determined by $ M $( or $ r $).
	\end{remark}
	
	In conclusion, we summarize these achievable computation rates and limiting rates in Table \ref{table rates}.

	\section{Proposed NOMA-CoMAC Scheme}
	\label{Design of NOMA-CoMAC}
	\begin{figure*}
		\centering
		\subfloat[Sub-Function Process and Superposition Process.]{
			\includegraphics[width=0.8\linewidth]{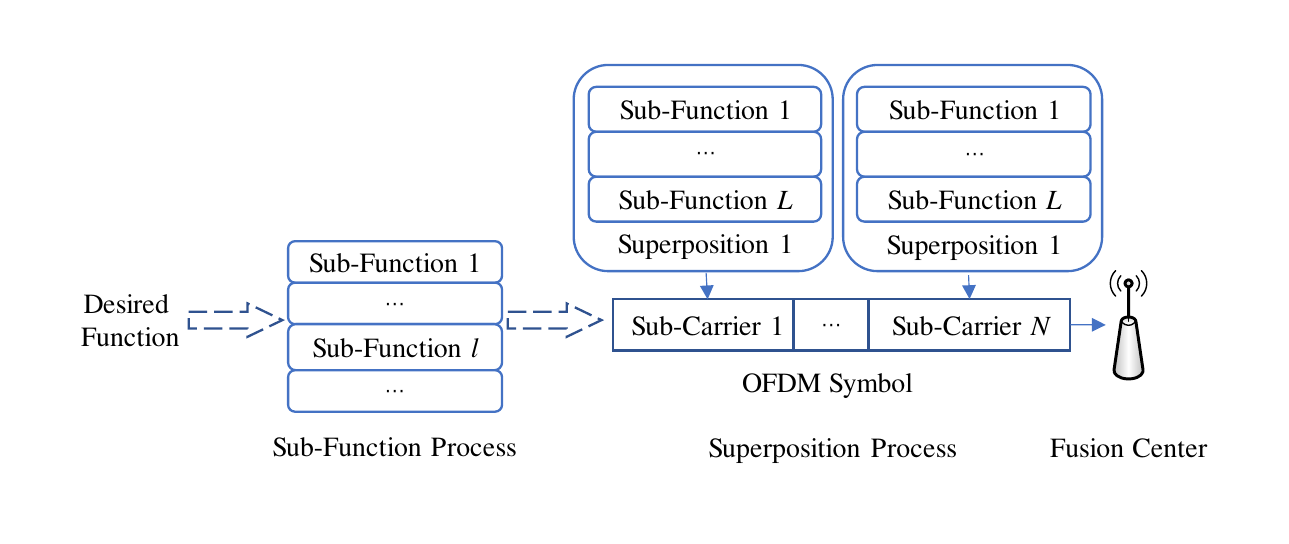}
			\label{fig:noma-comac-a}
		}
		\vfil
		\subfloat[SIC Process and Reconstruction Process.]{
			\includegraphics[width=0.9\linewidth]{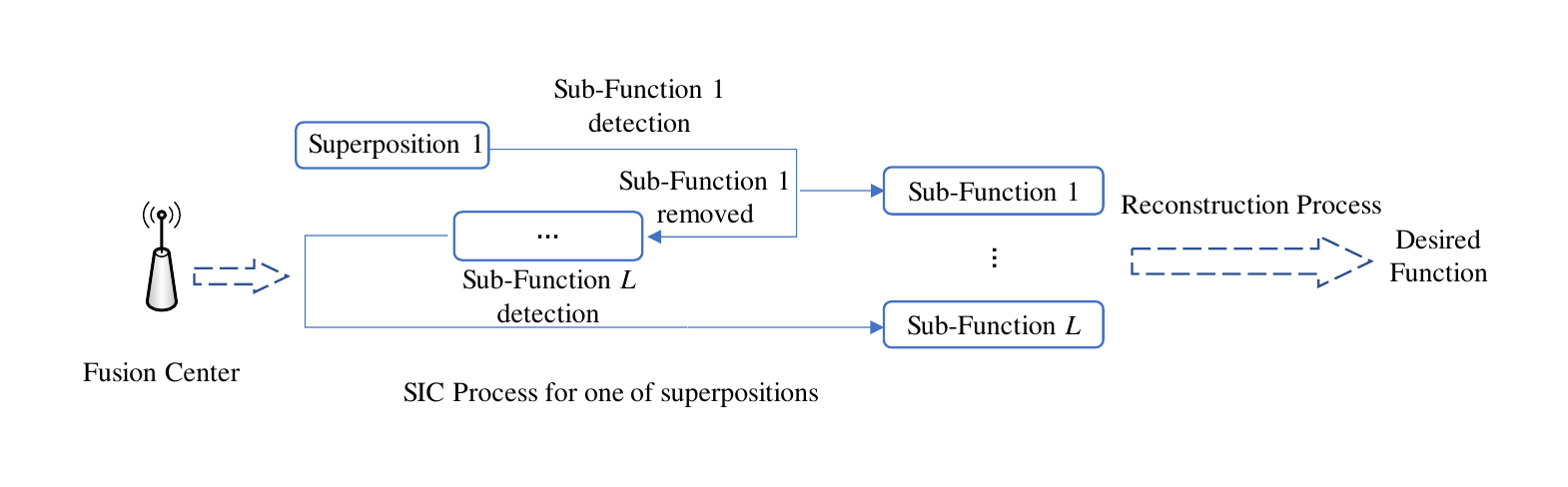}
			\label{fig:noma-comac-b}
		}
		\caption{Design of NOMA-CoMAC}
		\label{fig:omf-noma}
	\end{figure*}
	\label{Opportunistic Multi-Function NOMA for OFDM-based System}
	
	\subsection{Proposed Scheme}\label{OMF-NOMA Scheme}
	As mentioned in Section \ref{Previous Works}, NB-CoMAC only considered the flat fading channel. In order to improve the computation rate and deal with frequency selective fading, WB-CoMAC was proposed. These conventional CoMAC schemes transmit only one function (or sub-function) in each RB resulting in low spectrum utilization efficiency, hence applying NOMA design into CoMAC is a way to improve the computation rate through multiple sub-functions superposition.
	
	As shown in Fig.~\ref{fig:omf-noma}, we provide a simplified description on the proposed scheme in a hybrid OFDM-NOMA system. 
	
	\begin{itemize}
		\item \emph{Sub-Function Process.} In each sub-carrier, we sort all the nodes depending on the corresponding channel gains. Then, every $ M $ nodes in such an order computes a function which is regarded as a sub-function in Fig.~\subref*{fig:noma-comac-a}. Referring to Definition \ref{Partitioning Function}, $ \tau $ denotes the set whose elements belong to these indexes of $ M $ nodes to compute a sub-function $ f\left(\left\lbrace s_i[j]\right\rbrace_{i\in\tau} \right) $. Then, let the set
		\begin{equation}\label{key}
			\mathcal{S}=\left\lbrace \tau\subseteq[1:K]:\left|\tau \right|=M \right\rbrace 
		\end{equation}
		include all the possible sub-functions\footnote{For easy presentation, we use the element $ \tau\in\mathcal{S} $ stands for the sub-function $ f\left(\left\lbrace s_i[j]\right\rbrace_{i\in\tau} \right) $ which is computed by these nodes in $ \tau $.}, and the cardinality of $ \mathcal{S} $ is $ \left| \mathcal{S}\right|=\binom{K}{M} $.
		\item \emph{Superposition Process.} As shown in Fig.~\subref*{fig:noma-comac-a}, let the worst channel gain in the sub-function stand for the equivalent channel gain of the sub-function. Then, we sort all the sub-functions in each sub-carrier according to these equivalent channel gains. Only the first $ L $ sub-functions are chosen to be simultaneously transmitted, which is seen as a superposition. Then, one possible superposition can be defined as 
		\begin{equation}\label{key}
			\delta=\left\lbrace \tau_1,\cdots,\tau_L:\tau_u\cap\tau_v=\emptyset, \left|\delta \right|=L \right\rbrace,
		\end{equation}
		where $ u\neq v $, $ u,v\in[1:L] $ and sub-functions $ \tau_u, \tau_v \in \mathcal{S} $. All the possible superpositions are in a set 
		\begin{equation}\label{key}
			\mathcal{H}=\left\lbrace\delta\in\mathcal{S}: \left| \mathcal{H}\right|=\prod_{l=0}^{L-1}\binom{K-Ml}{M}  \right\rbrace.
		\end{equation}
		\item \emph{SIC Process.} As shown in Fig.~\subref*{fig:noma-comac-b}, all the OFDM symbols are received at the fusion center. Each sub-carrier contains a superposition with $ L $ sub-functions. Through SIC given in \cite{nazer2011compute}, we can obtain all the sub-functions.
		\item \emph{Reconstruction Process.} As mentioned in Definition \ref{Partitioning Function}, all the sub-functions need to be reconstructed at the fusion center. The set
		\begin{equation}\label{key}
	\begin{split}
		\mathcal{X}=\left\lbrace\varphi=\left\lbrace\delta_1,\delta_2,\cdots,\delta_{D} \right\rbrace:\delta_u\cap\delta_v=\emptyset,\vphantom{\left.\bigcup_{u=1}^{D}\delta_u=[1:K],\left| \varphi\right|=D=\frac{B}{L}\right\rbrace}\right.\left.\bigcup_{u=1}^{D}\delta_u=[1:K],\left| \varphi\right|=D=\frac{B}{L}\right\rbrace
	\end{split}
\end{equation}
		contains all the possible combinations whose element $ \varphi $ can reconstruct a whole desired function, and the cardinality of $ \mathcal{X} $ is $ \left| \mathcal{X}\right|=\prod_{l=0}^{L-1}\binom{ML-Ml}{M}\prod_{d=0}^{D-1}\binom{K-Md}{ML} $. After all the sub-functions are collected in Fig.~\subref*{fig:noma-comac-b}, we can recover the desired functions by using the relationship between the sub-functions and the desired functions.
	\end{itemize}

	\subsection{Computation Rate of NOMA-CoMAC}
	As shown in Fig.~\ref{fig:omf-noma}, the desired function is divided into $ B\in \mathbb{N} $ parts, each part can be regarded as a sub-function which is computed at fusion center individually. In each sub-carrier, $ L $ sub-functions are chosen to be transmitted. Based on those definitions in the previous sub-section, we use the following parts to derive the computation rate step by step.
	
	\textbf{\emph{Rate of Sub-Function $\tau$}}. As mentioned in Fig.~\subref*{fig:noma-comac-a}, $ L $ sub-functions are chosen for the $ g $-th sub-carrier at the $ m $-th OFDM symbol to be transmitted. The $ l $-th sub-function is computed  by $ M $ nodes whose indexes are in the set $ \left\lbrace\mathcal{I}_{u}^g[m] \right\rbrace_{u\in\mathcal{M}_l}   $ at the fusion center. We assume the bandwidth of the hybrid OFDM-NOMA system with $ N $ sub-carriers is the same as the mentioned conventional CoMAC system. The bandwidth that each sub-carrier owns is  $ \frac{1}{N} $ of the total bandwidth. Then, the computation rate of the $ l $-th sub-function in a sub-carrier at the $ m $-th OFDM symbol can be given as follows.
	
	\begin{lemma}[Computation Rate of a Sub-Function]\label{Rate of Multi-Function Superposition}
		With the limit of large $ n $ and $ L $ sub-functions in the $ g $-th sub-carrier, the instantaneous computation rate of the $ l $-th sub-function at the $ m $-th OFDM symbol with AWGN whose variance is $ \frac{1}{N} $ can be express as 
		\begin{equation}\label{l-th sub-function rate}
			\resizebox{!}{!}{
				$
				R_{l,g}[m]=\dfrac{1}{N}\mathsf{C}^{+}\left( \dfrac{N}{M}+\dfrac{\displaystyle N\min_{u\in\mathcal{M}_l}\left[ |h_{\mathcal{I}_u^g[m],g}[m]|^2P_{\mathcal{I}_u^g[m],g}[m]\right] }{1+N\displaystyle\sum_{j=l+1}^{L}\displaystyle\min_{u\in\mathcal{M}_j}\left[ |h_{\mathcal{I}_u^g[m],g}[m]|^2P_{\mathcal{I}_u^g[m],g}[m]\right] }\right),
				$
			}
		\end{equation}
		where $ |h_{\mathcal{I}_u^g[m],g}[m]|^2 $ is the channel gain of the $ g $-th sub-carrier for the node $ \mathcal{I}_u^g[m] $, $ P_{\mathcal{I}_u^g[m],g}[m] $ is the power allocated to the $ \mathcal{I}_u^g[m] $-th node in the $ g $-th sub-carrier and $ \mathcal{M}_l $ is the set including the index of the chosen nodes to compute the $ l $-th sub-function (See Eq.~\eqref{General Rate}).
	\end{lemma}
	
	\begin{IEEEproof}
		As demonstrated in \cite{nazer2011compute}, CoMAC can subtract part of the contribution from the channel observation to compute several functions at the fusion center based on successive cancellation. Let $ \bm{\mathrm{h}} $ denote the channel vector and $ \bm{\mathrm{a}}_l $ denote the coefficient vector to compute the $ l $-th function. From \cite[Theorem 12]{nazer2011compute}, the computation rate of the $ l $-th function from the channel observation with a noise variance of $ \sigma_{Z}^2 $  can be express as 
		\begin{equation}\label{key}
			R_{l}=\mathsf{C}^{+}\left(\dfrac{P}{|\alpha_l|^2+P\|\alpha_l\bm{\mathrm{h}}-\sum_{j=1}^{l}\bm{\mathrm{a}}_j\|^2 } \right),
		\end{equation}
		where $ \alpha_l $ is the scalar parameter to move the channel coefficients closer to the $ l $-th desired function.
		By giving the optimal $ \alpha_l $ following \cite[Remark 11]{nazer2011compute}, $ \frac{1}{N} $ of the noise variance and the $ i $-th element of the coefficient vector
		\begin{equation}\label{key}
			\bm{\mathrm{a}}_l[i]=\left\lbrace 
			\begin{aligned}
				\bm{\mathrm{h}}[i]\quad&i\in\left\lbrace\mathcal{I}_{u}^g[m] \right\rbrace_{u\in\mathcal{M}_l}\\
				0\quad&\mathrm{otherwise}
			\end{aligned},\right. 
		\end{equation}
		the computation rate of the $ l $-th sub-function in single sub-carrier can be given as
		\begin{equation}\label{naive single sub-carrier rate}
			R_{l}=\mathsf{C}^{+}\left( \dfrac{N}{M}+\dfrac{NP\|\bm{\mathrm{a}}_l\|^2 }{1+NP\sum_{j=i+1}^{L}\|\bm{\mathrm{a}}_j\|^2 }\right).
		\end{equation}
		
		Then combining Eq.~\eqref{naive single sub-carrier rate} with \cite[Theorem 3]{jeon2014computation}, the computation rate considering fading channel and power control at the $ t $-th time slot is further expressed as
		\begin{equation}\label{time domain rate for single sub-carrier}
			\resizebox{!}{!}{
				$
				R_{l}[t]=\mathsf{C}^{+}\left( \dfrac{N}{M}+\dfrac{\displaystyle N\min_{u\in\mathcal{M}_l}\left[ |h_{\mathcal{I}_u[t]}[t]|^2P_{\mathcal{I}_u[t]}[t]\right] }{1+N\sum_{j=l+1}^{L}\displaystyle\min_{u\in\mathcal{M}_j}\left[ |h_{\mathcal{I}_u[t]}[t]|^2P_{\mathcal{I}_u[t]}[t]\right] }\right).
				$
			}
		\end{equation}
		
		Since the propagation time of a sub-carrier symbol in OFDM needs $ N $ time slots as mentioned in \cite[Lemma 1]{wu2018computation}, $ R_{l,g}[m] $ for the $ l $-th sub-function in the $ g $-th sub-carrier at the $ m $-th OFDM symbol is $ \frac{1}{N} $ of $ R_{l}[t] $.  In conclusion, Lemma \ref{Rate of Multi-Function Superposition} has been proved.		
	\end{IEEEproof}
	
	\textbf{\emph{Rate of Superposition $\delta$}}. Compared with conventional CoMAC schemes, each sub-function in the same sub-carrier has to face inter-function interference in NOMA-CoMAC, which causes the different computation rates of different sub-functions in the same sub-carrier. Lemma \ref{Rate of Multi-Function Superposition} demonstrates the rate of the $ l $-th function in the $ g $-th sub-carrier is a part of the sum rate of the superposition of the $ L $ sub-functions in single sub-carrier. From Fig.~\subref*{fig:noma-comac-a}, it shows that we need to determine the sum rate of all the $ L $ sub-functions in a superposition since those sub-functions are transmitted as a whole.
	
	\begin{lemma}[Computation Rate of a Superposition]\label{Ex Rate of Multi-Function Superposition}
		As the limit of large $ n $, the instantaneous computation rate of the superposition of $ L $ sub-functions in the $ g $-th sub-carrier at the $ m $-th OFDM symbol with AWGN whose variance is $ \frac{1}{N} $ can be express as 
		\begin{align}
			R^{\delta}_{\varphi,g}[m]=\min_{l\in[1:L]}R_{l,g}[m]
		\end{align}
		
	\end{lemma}
	
	\begin{IEEEproof}
		The rate of the superposition of $ L $ sub-functions is determined by the minimum $ R_{l,g}[m] $ for all $ l\in[1:L] $, since each sub-function is a part of the original desired function and the desired function can be reconstructed if and only if all parts have been received at the fusion center. 
	\end{IEEEproof}
	
	\textbf{\emph{Rate of Combination $\varphi$}}. In the hybrid OFDM-NOMA system with $ N $ sub-carriers, the number of OFDM symbols $ T_s $ is at least $   \frac{n}{N}  $\footnote{In order to simplify the derivation, the number of OFDM symbols in Eq.~\eqref{OFDM-NOMA model} is given as $ \frac{n}{N} $ instead.}. It also implies that the number of all the sub-carriers during $ T_s $ OFDM symbols is $ n $, and each sub-carrier serves one superposition $ \delta\in\mathcal{H} $. We define a set $ \mathcal{M}_{\varphi} $ including those sub-carriers that serve the combination $ \varphi $ and a set $ \mathcal{M}_{\varphi}^{\delta} $ containing the sub-carriers that serve the specific superposition $ \delta $ in the combination $ \varphi $. Since the superpositions and the combinations in practice are random depending on channel realizations, it causes that $ |\mathcal{M}_{\varphi}| $ and $ |\mathcal{M}_{\varphi}^{\delta}|$ are stochastic. As the limit of large $ n $, the set $ \mathcal{M}_{\varphi}^{\delta} $ and $ \mathcal{M}_{\varphi} $  contain, respectively, $ |\mathcal{M}_{\varphi}^{\delta}|=\frac{n}{D|\mathcal{X}|} $ sub-carriers and $ |\mathcal{M}_{\varphi}|=D|\mathcal{M}_{\varphi}^{\delta}| $ sub-carriers referring to \cite[Lemma 2]{wu2018computation}. Then, the transmission of the specific superposition  $ \delta $ in the combination $\varphi$ totally occupies $ T^{\delta}_{\varphi}=\lceil  \frac{|\mathcal{M}_{\varphi}^{\delta}|}{N} \rceil $ OFDM symbols\footnote{Although these superpositions are sent separately in different sub-carrier in different OFMD symbol, we can consider that they are transmitted centrally when obtaining the achievable rate.}.		
	
	\begin{lemma}[Computation Rate of a Combination]\label{Rate of Combination}
		The average rate for computing those sub-functions in the combination $\varphi$ during $ T_{\varphi}=DT^{\delta}_{\varphi} $ OFDM symbols can be given as
		
		\begin{equation}\label{Avarage rate for L sub-functions in OFDM}
			R_{\varphi}=\dfrac{1}{T_{\varphi}}\sum_{m=1}^{T^{\delta}_{\varphi}}\dfrac{1}{N}\sum_{g=1}^{N}R^{\delta}_{\varphi,g}[m].
		\end{equation}
	\end{lemma}
	
	\begin{IEEEproof}
		According to Eqs.~\eqref{OFDM-NOMA model} and \eqref{OFDM-NOMA model of sub-carrier}, the received signal in the $ g $-th sub-carrier for the superposition $ \delta $ based on the combination $ \varphi $ can be given as
		\begin{align}
			y_{\varphi,g}^{\delta}[m]=&\sum_{\tau_l\in\delta}\sum_{i\in\tau_l}v_{i,\varphi,g}^{\tau_l}[m]x_{i,\varphi,g}^{\tau_l}[m]h^{\tau_l}_{i,\varphi,g}[m]+w_{i,\varphi,g}^{\tau_l}[m]\\
			=&\sum_{\tau_l\in\delta}\sum_{i\in\tau_l}x_{i,\varphi,g}^{\tau_l}[m]h^{'\tau_l}_{i,\varphi,g}[m]+w_{i,\varphi,g}^{\tau_l}[m]\label{OMF-NOMA in single sub-carrier},
		\end{align}
		where $ m=[1:T_{\varphi}^{\delta}] $,$ \tau_l $ contains the indexes of the $ l $-th sub-function in the superposition $ \delta $ and $ h^{'\tau_l}_{i,\varphi,g}[m]=|h^{\tau_l}_{i,\varphi,g}[m]|\sqrt{P_{i,\varphi,g}^{\tau_l}[m]} $ is equivalent channel.

		Depending on the conclusions of Lemmas \ref{Rate of Multi-Function Superposition},~\ref{Ex Rate of Multi-Function Superposition} and Eq.~\eqref{OMF-NOMA in single sub-carrier}, we can obtain the average rate for computing the superposition $ \delta $ in the combination $\varphi$ during $ T^{\delta}_{\varphi} $ OFDM symbols 
		\begin{equation}
			R_{\varphi}^{\delta}=\dfrac{1}{T^{\delta}_{\varphi}}\sum_{m=1}^{T^{\delta}_{\varphi}}\dfrac{1}{N}\sum_{g=1}^{N}R_{\varphi,g}^{\delta}[m].
		\end{equation}
		
		Then, the average rate for computing those sub-function in the combination $\varphi$ is expressed as
		\begin{align}
			R_{\varphi}
			\stackrel{(a)}{=}\dfrac{1}{D}\min_{\delta\in\varphi}R_{\varphi}^{\delta}\stackrel{(b)}{=}\dfrac{1}{D}R_{\varphi}^{\delta},
		\end{align}
		where condition $ (a) $ follows because of the similar result to Lemma \ref{Ex Rate of Multi-Function Superposition} and condition $ (b) $ follows as each average rate $ R_{\varphi}^{\delta} $ approaches the same value with the limit of large $ n $.
	\end{IEEEproof}

	With the help of Eq.~\eqref{Avarage rate for L sub-functions in OFDM} and the length of the transmitted vector  $ |\mathcal{M}_{\varphi}| $, the length of the data vector is $ U_{\varphi}=\dfrac{R_{\varphi}|\mathcal{M}_{\varphi}|}{H(f(\bm{\mathrm{s_r}}))} $ as the same as the number of the desired function values reconstructed by combination $ \varphi $ (See Definitions \ref{Encoding & Decoding} and \ref{Computation rate}). $ U_{\varphi} $ is only the part of all the values of desired function $ T_d $, and the exact number of desired function values for all $ \varphi\in\mathcal{X} $ during $ T_s $ OFDM symbols is 
	\begin{equation}\label{length of T_d}
		T_d=\sum_{\varphi\in\mathcal{X}}U_{\varphi}.
	\end{equation}

	Finally, with the help of Lemmas \ref{Rate of Multi-Function Superposition}, \ref{Ex Rate of Multi-Function Superposition} and \ref{Rate of Combination}, the computation rate of NOMA-CoMAC based on Definition \ref{Computation rate} can be given as
	\begin{align}
		R&=\lim\limits_{n\rightarrow \infty}\dfrac{T_d}{n}H(f(\bm{\mathrm{s_r}}))\nonumber\\
		&=\lim\limits_{n\rightarrow \infty}\dfrac{\displaystyle\sum_{\varphi\in\mathcal{X}}\dfrac{R_{\varphi}|\mathcal{M}_{\varphi}|}{H(f(\bm{\mathrm{s_r}}))}}{n}H(f(\bm{\mathrm{s_r}}))\nonumber\\
		&=\dfrac{ML}{KN}\dfrac{1}{T_{s} }\sum_{m=1}^{T_{s}}\sum_{g=1}^{N}R^{\delta}_{\varphi,g}[m].
	\end{align}
	In conclusion, the rate in Theorem \ref{OMF-NOMA rate} is achievable as $ n $ increases.

	\section{Performance of Proposed NOMA-CoMAC Scheme}
	\label{Power Control for NOMA-CoMAC}
	\label{Intra-Block Power Control}
	In this section, we derive the achievable computation rate of the proposed NOMA-CoMAC with average power control based on the general rate in Theorem \ref{OMF-NOMA rate}. We further analyze the outage performance and obtain the diversity order.

	
	%
	
	\subsection{Power Control} We consider an average power control method where the average power of each node in each sub-carrier is no more than $\frac{P}{N} $ for all $ g\in[1:N] $, i.e., $ \mathsf{E}\left[ P_{i,g}[m]\right] \le \frac{P}{N} $. Let $ P_{i,g}[m] $ represent the transmitted power in the $ g $-th sub-carrier at the $ m $-th OFDM symbol for the node $ i $, and it can be express as
	\begin{equation}\label{APowerControl}
		\resizebox{!}{!}{
			$
			P_{\mathcal{I}_{i}^g[m],g}[m]=\left\{
			\begin{aligned}
			&\beta_{l,g}[m]\dfrac{KP\dfrac{|h_{\mathcal{I}_{Ml}^g[m],g}[m]|^2}{|h_{\mathcal{I}_{i}^g[m],g}[m]|^2}}{\displaystyle N\sum_{l=1}^{L}\beta_{l,g}[m]\varpi_{l,g}}&i\in\mathcal{M}_l,\forall l\in[1:L] \\
			&0&\rm{otherwise}
			\end{aligned}
			\right.,
			$
		}
	\end{equation}
	where $\varpi_{l,g}=\mathsf{E}\left[\frac{|h_{\mathcal{I}_{Ml}^g[m],g}[m]|^2}{|h|^2} \right]=\mathsf{E}\left[\frac{|h_{\mathcal{I}_{Ml}^g,g}|^2}{|h|^2} \right]$ as a constant, $ \beta_{l,g}[m] $ can be regarded as the power factor to compute the $ l $-th sub-function in the $ g $-th sub-carrier and the detailed derivation is given in Appendix \ref{Proof of Corollary intra-block power alloaction}. By putting Eq.~\eqref{APowerControl} into Eq.~\eqref{l-th sub-function rate}, the rate of the $ l $-th sub-function $ R_{l,g}[m] $ in Lemma \ref{Rate of Multi-Function Superposition} can be rewritten as
	\begin{equation}\label{key}
		\resizebox{!}{!}{
			$
			R_{l,g}[m]=\dfrac{1}{N}\mathsf{C}^{+}\left( \dfrac{N}{M}+\dfrac{PK|h_{\mathcal{I}_{Ml}^g[m],g}[m]|^2\beta_{l,g}[m] }{M\sum_{l=1}^{L}\beta_{l,g}[m]\varpi_{l,g}+\sum_{j=l+1}^{L}PK|h_{\mathcal{I}_{Mj}^g[m],g}[m]|^2\beta_{j,g}[m]} \right).
			$
		}
	\end{equation}
	
	
	\subsection{Problem Formulation} We work on maximizing the instantaneous rate of each OFDM symbol to improve the ergodic rate, since the rate in Theorem \ref{OMF-NOMA rate} can be regarded as the mean of the instantaneous rate. Then we formulate the following optimization.
	
	\begin{problem}
		\label{optimization for intra-block power control}
		\begin{align}
			\mathop{\mathrm{maximize}}\limits_{\beta_{l,g}[m]}& \quad \sum_{g=1}^{N}\min_{l\in[1:L]}R_{l,g}[m] \nonumber \\
			s.t.&\quad \beta_{l,g}[m]\ge0\quad \forall l\in[1:L] \nonumber
		\end{align}
	\end{problem}
	
	Because the superposition transmission of too many sub-functions brings the difficulty of SIC at the fusion center and makes the mathematical analysis hard, we choose two sub-functions as a pair to be transmitted in single sub-carrier. By setting $ L=2 $ in Problem \ref{optimization for intra-block power control}, the relationship between $ \beta_{1,g}[m] $ and $ \beta_{2,g}[m] $ can be obtained as
	\begin{equation}\label{the relationship between factors}
	\resizebox{\linewidth}{!}{$
		\begin{aligned}
		\beta_{1,g}[m]=\left(\sqrt {
			\left( \Upsilon[m]\,{ \varpi_{2,g}}+{ \varpi_{1,g}} \right) ^{2}+4
			\,{ P\frac{K}{M}}\,{ |h_{\mathcal{I}_{M}^g[m],g}[m]|^2}{ \varpi_{1,g}} }\right.\left.\vphantom{\sqrt {
				\left( \Upsilon\,{ \varpi_{2,g}}+{ \varpi_{1,g}} \right) ^{2}+4
				\,{ P\frac{K}{M}}\,{ |h_{\mathcal{I}_{M}^g[m],g}[m]|^2}{ \varpi_{1,g}} }}+{ \varpi_{1,g}}-\Upsilon[m]\,{ \varpi_{2,g}} \right)\dfrac{\beta_{2,g}[m]}{2{\Upsilon[m]\,{ \varpi_{1,g}}}},
		\end{aligned}
		$}
\end{equation}
	where $ \Upsilon[m]=\frac{{ |h_{\mathcal{I}_{M}^g[m],g}[m]|^2}}{{ |h_{\mathcal{I}_{2M}^g[m],g}[m]|^2}} $ and the detailed derivation is given in Appendix \ref{Proof of Corollary intra-block power alloaction}.
	
	Therefore, based on Theorem \ref{OMF-NOMA rate} and the increasing $ n $, the ergodic computation rate of NOMA-CoMAC with a sub-function pair
	\begin{equation}
		\resizebox{!}{!}{$
			\begin{aligned}
			R\stackrel{(a)}{=}\dfrac{2M}{KNT_s}\displaystyle\sum_{m=1}^{T_s}\left[\sum_{g=1}^{N}R_{1,g}[m]\right]\stackrel{(b)}{=}\dfrac{2M}{KNT_s}\sum_{m=1}^{T_s}\left[\sum_{g=1}^{N}R_{2,g}[m]\right]
			\end{aligned}
			$}
	\end{equation}
	is achievable, where the conditions $ (a) $ and $ (b) $ follow because $ R_{1,g}[m]=R_{2,g}[m] $ with the given optimal power allocation factors $ \beta_{1,g}[m] $ and $ \beta_{2,g}[m] $ in Eq.~\eqref{the relationship between factors}. As a result, Corollary \ref{intra-block power alloaction} has been proved.
	
	Note that the power allocated to each node for computing the corresponding sub-function in Eq.~\eqref{APowerControl} is a closed-form expression with the given relationship between $ \beta_{1,g}[m] $ and $ \beta_{2,g}[m] $ in Eq.~\eqref{the relationship between factors}. It implies that the allocated power can be calculated directly with low computational complexity. Hence, the method of power control is suitable to be deployed in a system that requires low latency.
	
	\subsection{Outage Probability and Diversity Order} It is worth obtaining the outage probability and the diversity order for NOMA-CoMAC system, which can help us to analyze the performance theoretically. An exact expression for the outage probability and the diversity order are provided in the following corollary.
	
	\begin{corollary}[Diversity Order of NOMA-CoMAC]\label{NOMA gain}
		The outage probability for NOMA-CoMAC with a sub-function pair transmission in the OFDM-based system can be given as
	\begin{equation}\label{outage probability}
	\resizebox{!}{!}{$
		\begin{split}
		P_{out}=&\left\lbrace 1-\Theta\sum_{u_1=0}^{M-1}\sum_{u_2=0}^{K-2M}{{M-1}\choose{u_1}}{{K-2M}\choose{u_2}}(-1)^{K-2M+u_1-u_2}\right.\\
		&\times\left[ \int_{\frac{\xi}{P}}^{\infty}\int_{\phi_1}^{\phi_2}\mathrm{Ex}_1(x)\mathrm{d}x\mathrm{Ex}_2(y)\mathrm{d}y\right.\left.\left.+\int_{\frac{\varepsilon_{R^T}}{NP\beta_{2M}}}^{\infty}\int_{\phi_3}^{\infty}\mathrm{Ex}_1(x)\mathrm{d}x\mathrm{Ex}_2(y)\mathrm{d}y\right] \right\rbrace ^{N},
		\end{split}
		$}
\end{equation}
		where $ \Theta= \frac{K!}{(M-1)!(M-1)!(L-2M)!}$, $ \mathrm{Ex}_1(x)=e^{-(M+u_1)x} $, $ \mathrm{Ex}_2(y)=e^{-(K-M-u_1-u_2)y} $, $ \phi_1= \frac{\varepsilon_{R^T}}{N\beta_{1,g}P}+\frac{\varepsilon_{R^T}\beta_{2,g}}{\beta_{1,g}}|h_{\mathcal{I}_{2M}^g,g}|^2 $, $\phi_2= \frac{\beta_{2,g}}{\beta_{1,g}}|h_{\mathcal{I}_{2M}^g,g}|^2+NP\frac{\beta_{2,g}}{\beta_{1,g}}|h_{\mathcal{I}_{2M}^g,g}|^4 $, $\phi_3= \frac{\beta_{2,g}}{\beta_{1,g}}|h_{\mathcal{I}_{2M}^g,g}|^2+NP\frac{\beta_{2,g}}{\beta_{1,g}}|h_{\mathcal{I}_{2M}^g,g}|^4 $, $ \xi=\frac{\varepsilon_{R^T}-1+\sqrt{\frac{4\varepsilon_{R^T}}{\beta_{2,g}}+(\varepsilon_{R^T}-1)^2}}{2N} $ and $ \varepsilon_{R^T}=2^{\frac{R^TKN}{2M}}-\frac{N}{M} $.
		
		The diversity order achieved by NOMA-CoMAC with a sub-function pair transmission is given by
		\begin{equation}\label{key}
			-\lim\limits_{P\to\infty}\dfrac{\log{P_{out}}}{\log{P}}=\underbrace{N}_{o_1}\cdot\underbrace{(K-2M)}_{o_2}
		\end{equation}
	\end{corollary}
	
	\begin{IEEEproof}
		See Appendix \ref{Proof of Corollary NOMA gain}.
	\end{IEEEproof}
	
	Corollary \ref{NOMA gain} illustrates that the diversity order in our proposed NOMA-CoMAC with a sub-function pair transmission consists of two parts.
	The first part $ o_1 $ is achievable benefiting from the OFDM design, and the superposition of the first $ L=2 $ sub-functions brings the gain as the second part $ o_2 $. Furthermore, the part $ o_2 $ shows that the node with the worst channel gain in these two sub-functions plays a dominant role.

	\section{Numerical Results and Discussion}
	\label{Simulation Results and Discussions}
	In this section, numerical results of computation rate based on NOMA-CoMAC are provided and compared with the conventional CoMAC schemes. Besides, the outage performance of NOMA-CoMAC is demonstrated, and the accuracy of the derived analytical results are verified through Monte Carlo simulation.
	\subsection{Computation Rate}
	\begin{figure}
		\centering
		\includegraphics[width=0.7\linewidth]{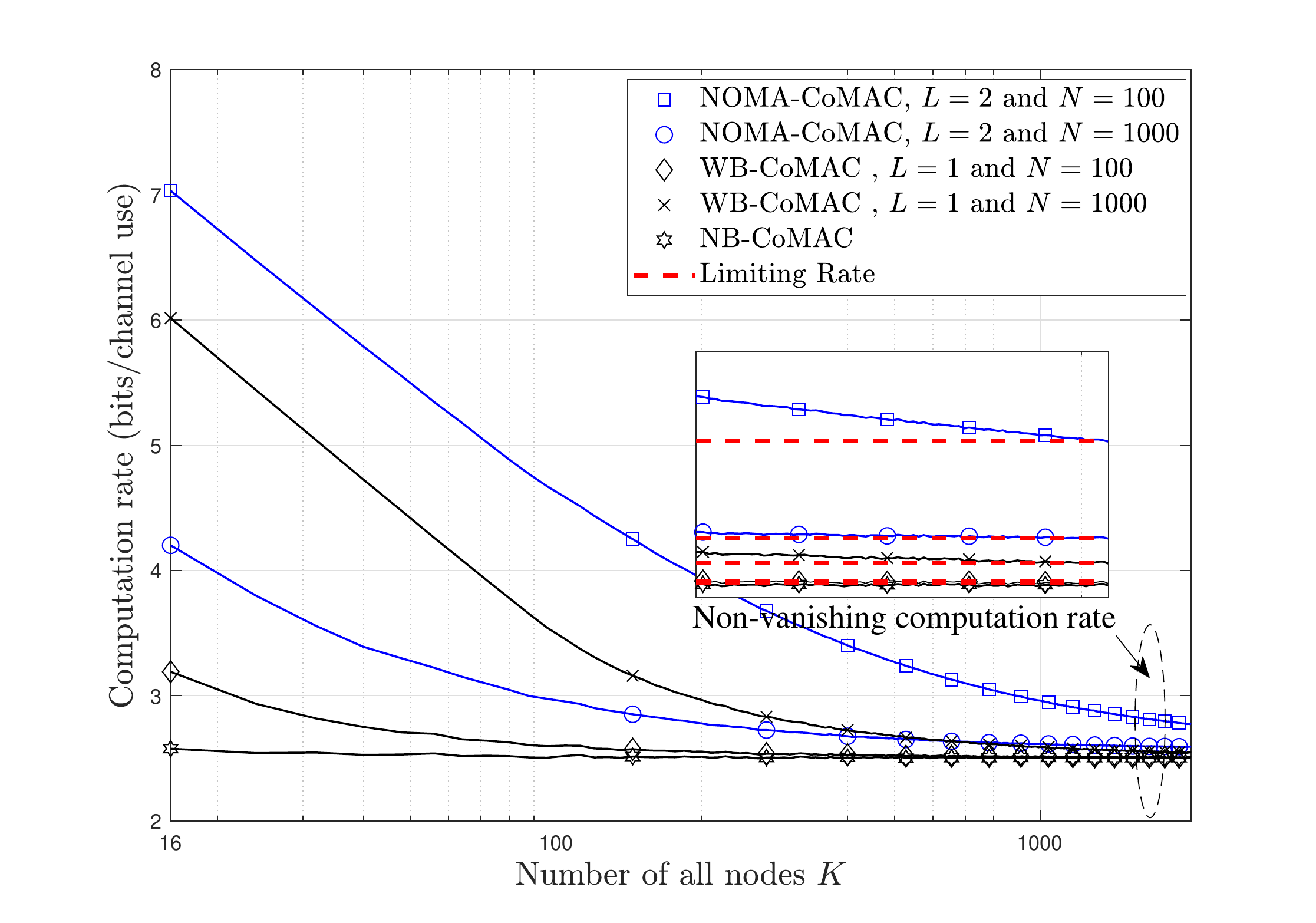}
		\caption{Comparison between NOMA-CoMAC and conventional CoMAC schemes versus the number of all nodes $ K $, ${\rm{SNR}} =10 $ dB}
		\label{fig:cs2}
	\end{figure}
	\begin{figure}
		\centering
		\includegraphics[width=0.7\linewidth]{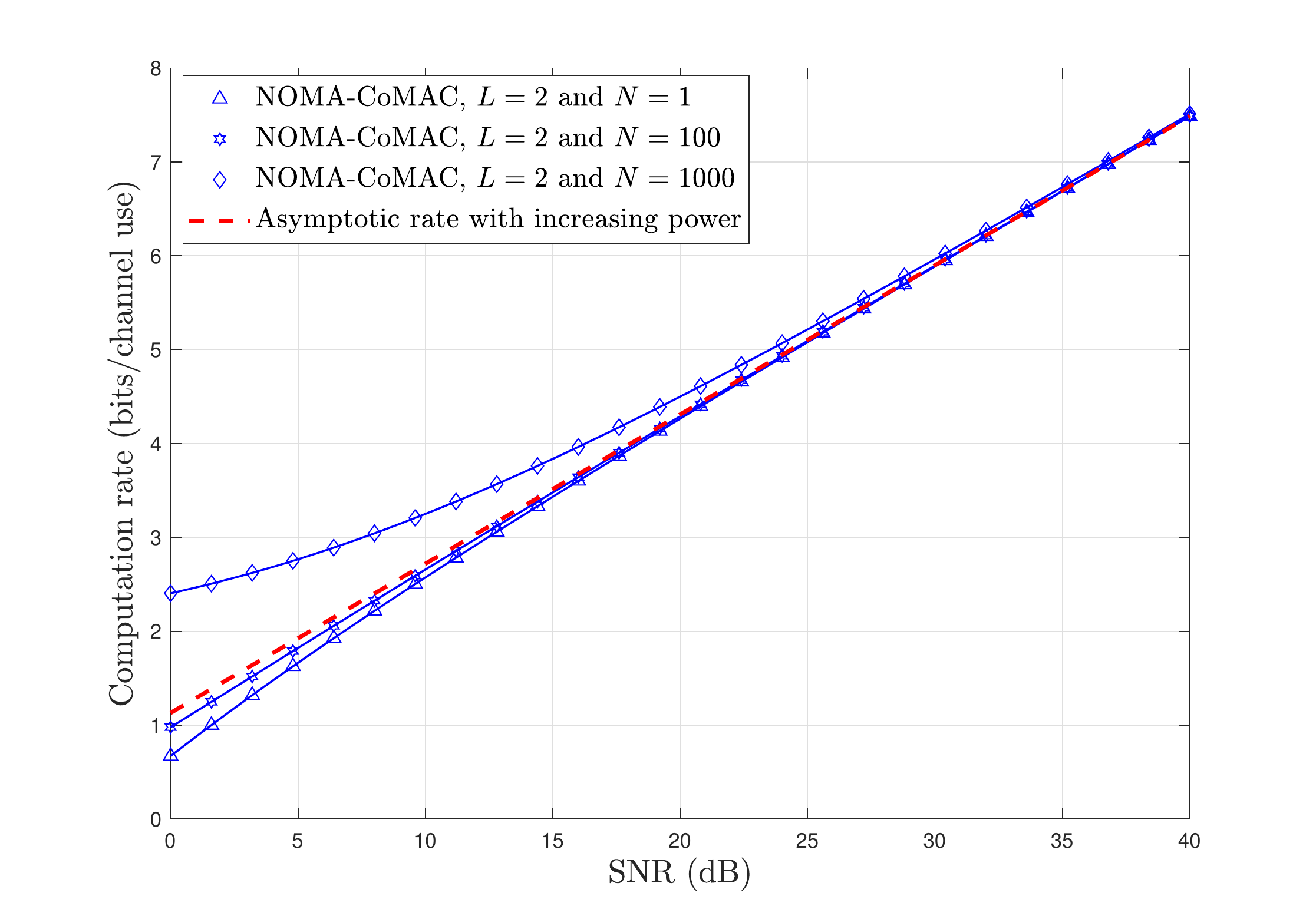}
		\caption{Computation rates of NOMA-CoMAC versus $ {\rm{SNR}} $ and the number of sub-carriers $ N $}
		\label{fig:cs6}
	\end{figure}
	
	We compare the conventional CoMAC schemes, i.e., NB-CoMAC and WB-CoMAC, with NOMA-CoMAC in Fig.~\ref{fig:cs2}. We see that NOMA-CoMAC can improve the computation rate over the conventional CoMAC schemes as the number of nodes $ K $ is small. When $ K $ increases, all the rates shown in Fig.~\ref{fig:cs2} are decreasing but keep a non-vanishing rate. Compared with the conventional CoMAC schemes, the proposed NOMA-CoMAC can provide an improved non-vanishing rate. It also verifies the limiting rates in Corollary \ref{Limiting Rate of OMF-NOMA}, and implies that the exact non-vanishing rates as the lower bounds can be obtained without any simulation using Corollary \ref{Limiting Rate of OMF-NOMA}.

	The relationship between the computation rate of NOMA-CoMAC and $ {\rm{SNR}} $ is shown in Fig.~\ref{fig:cs6}. When $ {\rm{SNR}} $ increases, all the three rates with different $ N $ increase. When $ {\rm{SNR}} $ is small, the number of sub-carriers plays the main role to improve the rate since the increasing $ N $ can reduce the equivalent noise power. However, when $ {\rm{SNR}} $ increases, the contribution of $ N $ becomes smaller, and can be ignored compared to $ {\rm{SNR}} $. Hence, all the three rates are asymptotically equal. From Corollary \ref{Limiting Rate of OMF-NOMA}, the asymptotic rate with increasing $ {\rm{SNR}} $ can be given as $ 2r{\mathsf{C}^ + }( {{P^{\frac{1}{2}}}\mathop \xi \nolimits_{1 - r}^{\frac{1}{2}} } ) $.

	\begin{figure}
		\centering
		\includegraphics[width=0.7\linewidth]{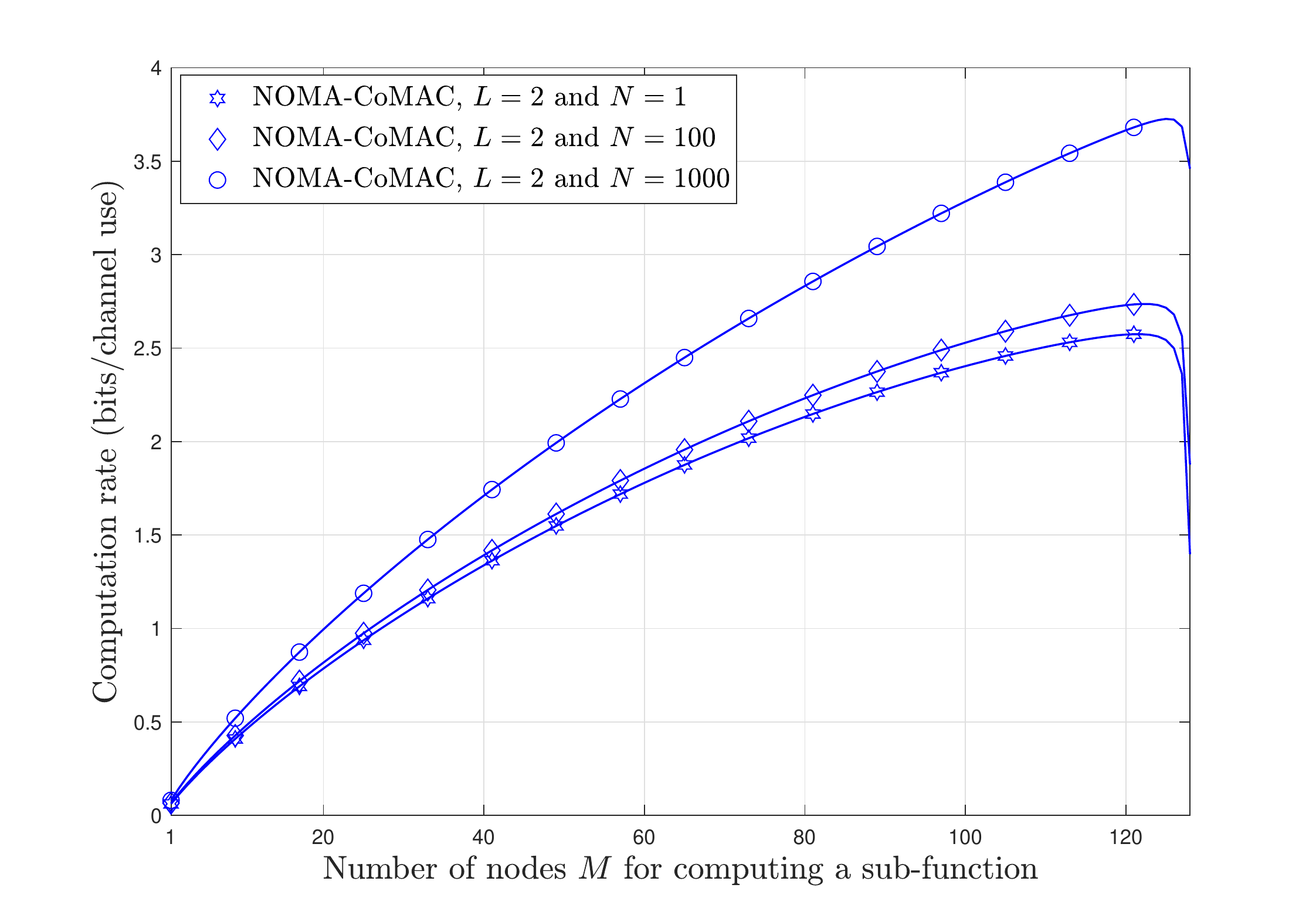}
		\caption{Computation rates of NOMA-CoMAC versus the number of chosen nodes $ M $, $ {\rm{SNR}} =10 $ dB}
		\label{fig:cs1}
	\end{figure}
	Fig.~\ref{fig:cs1} illustrates that the computation rate of NOMA-CoMAC referred to Corollary \ref{intra-block power alloaction} versus the increase of the number of chosen nodes $ M $. It shows that the rate in Eq.~\eqref{ex-rate} does not increase monotonically as $ M $ increases. The reason is that the rate gain $ \frac{2M}{KN} $ increases with $ M $, whereas the worse channel gain $ |h_{\mathcal{I}_{2M}^g[m],g}[m]|^2$ becomes vanishing as $ M $ increases. Hence, there is a trade-off between the rate gain and the worse channel gain. It also implies that there is an optimal $ M $ that achieves the maximum rate. In addition, with the increase of the number of sub-carriers $ N $, the rate can be improved because the equivalent noise power of lattice code decreases.
	\begin{figure}
		\centering
		\subfloat[the outage probabilities with different sub-function pairs.]{
			\includegraphics[width=0.6\linewidth]{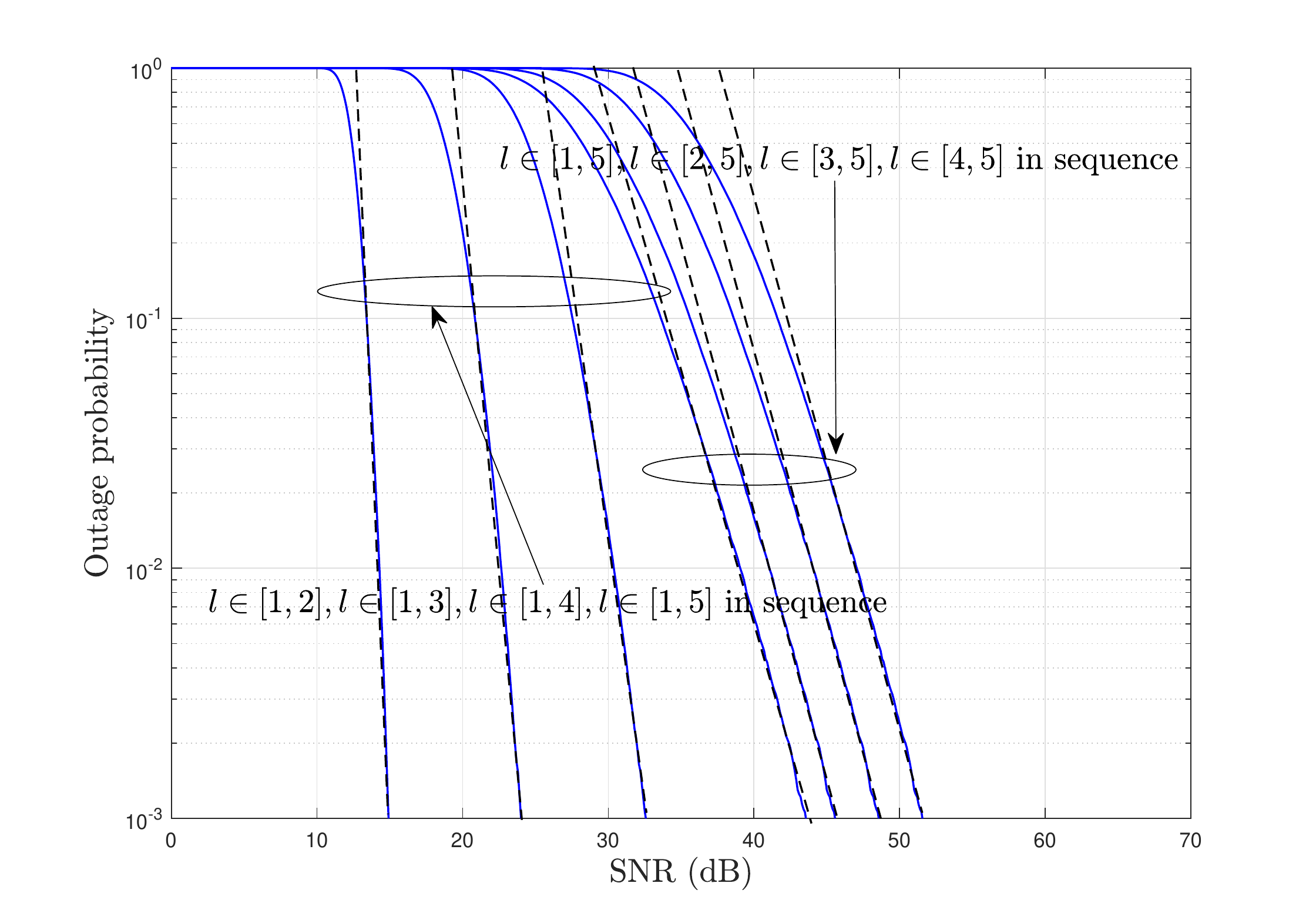}
			\label{fig:cs41}
		}
		
		\subfloat[the outage probabilities with different number of sub-carriers.]{
			\includegraphics[width=0.6\linewidth]{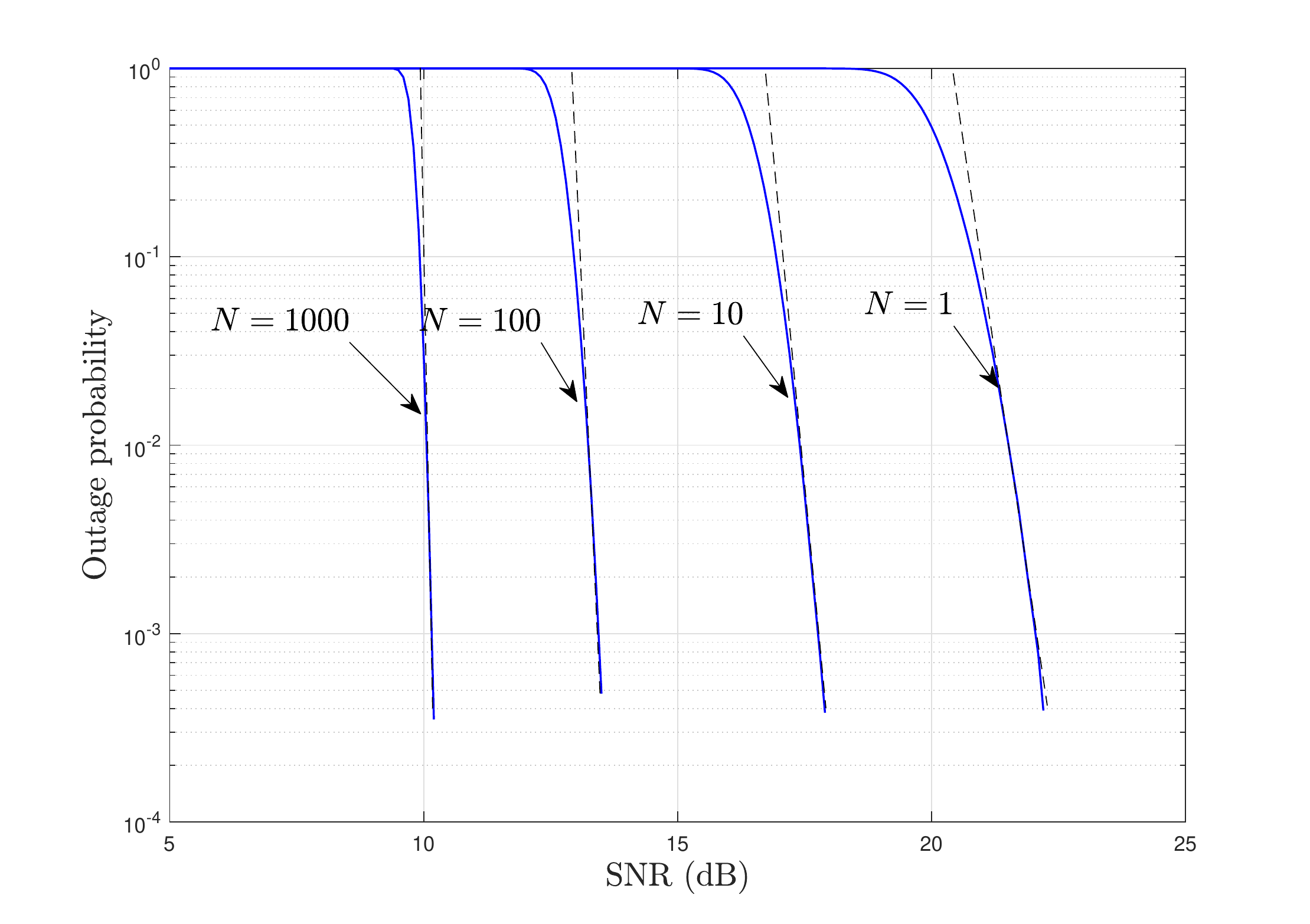}
			\label{fig:cs42}
		}
		\caption{Outage probabilities with different sub-function pairs and different number of sub-carriers.}
		\label{fig:cs4}
	\end{figure}
	
	\subsection{Outage Performance}
	
	As mentioned in Corollary \ref{intra-block power alloaction}, NOMA-CoMAC only chooses a sub-function pair to be computed in a single sub-carrier. Fig.~\subref*{fig:cs41} shows the outage performance of different sub-function pairs with the same first sub-function including $ l\in[1,2],l\in[1,3],l\in[1,4],l\in[1,5] $. One can observe that the outage performance of the sub-function pair $ l\in[1,2] $ is always superior to that of other sub-function pairs. Furthermore, the outage performance of different sub-function pairs with the same second sub-function including $ l\in[1,5],l\in[2,5],l\in[3,5],l\in[4,5] $ is also demonstrated in Fig.~\subref*{fig:cs41}. We can see that the outage probabilities for all five sub-function pairs have the same diversity order while the sub-function pair $ l\in[1,5] $ offers a
	constant performance gain over the others. As shown in Fig.~\subref*{fig:cs42}, both the diversity order and performance gain increase as the number of sub-carriers increases.

	Based on Fig.~\ref{fig:cs4}, we can see that the first function in the sub-function pair determines the performance gain. The diversity order depends on the second function of the sub-function pairs. Hence, the sub-function pair chosen in Corollary \ref{intra-block power alloaction} outperforms any other sub-function pair. Besides, the figures above verify Corollary \ref{NOMA gain} where the diversity order is only associated with the sub-function with poorer equivalent channel gain in the sub-function pair, and the diversity order increases as $ N $ increases.
	
	%
	%

	\section{Conclusion}
	\label{Conclusion}
	In order to provide extra improvement in spectrum efficiency, we have proposed a NOMA-CoMAC system through the division, superposition, SIC and reconstruction of the desired functions. Unlike those NOMA designs for information transmission, the proposed NOMA-CoMAC aims at computing functions over the MAC through superposing sub-functions. The expression of achievable computation rate has been derived based on nested lattice coding. It not only improves the spectrum efficiency, but also enhances the non-vanishing rate via sub-function superposition. Furthermore, we have considered some limiting cases to find more insights. As the number of nodes goes to infinity, we have obtained the exact expression of the limiting computation rate which characterizes the lower bound of the computation rate of NOMA-CoMAC. It can be used to evaluate the performance of the CoMAC system without time-consuming simulation. We have discussed the outage performance and analyzed the diversity order as the power goes to infinity. It shows the node with the worst channel gain among these sub-functions in each RB plays a dominant role.

	\appendices
	
	\section{Proof of Corollary \ref{intra-block power alloaction}}
	\label{Proof of Corollary intra-block power alloaction}
	
	We consider the average power control method where $ \mathsf{E}\left[ P_{i,g}[m]\right] \le \frac{P}{N} $. Let $ P_{i,g}[m] $ represent the transmitted power in the $ g $-th sub-carrier at the $ m $-th OFDM symbol for the node $ i $, which can be express as
	\begin{equation}\label{origin power for each user}
		\resizebox{!}{!}{$
			P_{\mathcal{I}_{i}^g[m],g}[m]=\left\lbrace 
			\begin{aligned}
			&c\beta_{l,g}[m]\dfrac{|h_{\mathcal{I}_{Ml}^g[m],g}[m]|^2}{|h_{\mathcal{I}_{i}^g[m],g}[m]|^2}&i\in\mathcal{M}_l,\forall l\in[1:L]\\
			&0&\rm{otherwise}
			\end{aligned}\right..
			$}
	\end{equation}
	
	In Eq.~\eqref{origin power for each user}, $ c $ is a constant which is needed to be solved, and $ \beta_{l,g}[m] $ can be regarded as the power factor to compute the $ l $-th sub-function in the $ g $-th sub-carrier. With the average power control, we have
	\begin{align}\label{key}
		\mathsf{E}[P_{i,g}[m]]&=\sum_{j=1}^{K}\Pr(i=\mathcal{I}_{j}^g[m])\mathsf{E}[P_{i,g}[m]|i=\mathcal{I}_{j}^g[m]] \nonumber\\
		&\stackrel{(a)}{=}c\frac{M}{K}\sum_{l=1}^{L}\beta_{l,g}[m]\underbrace{\mathsf{E}\left[\dfrac{|h_{\mathcal{I}_{Ml}^g[m],g}[m]|^2}{|h|^2} \right]}_{\varpi_{l,g}}=\dfrac{P}{N},
	\end{align}
	where condition $ (a) $ follows because the channel gains in different sub-carriers are i.i.d. random variables and $ h $ is used as a representative random variable without loss of generality. Then we can calculate the constant $ c $ and obtain Eq.~\eqref{APowerControl}.
	
	Assuming that the channel gain in each sub-carrier is i.i.d., and we further convert the $\rm{max-min} $ problem shown as Problem \ref{optimization for intra-block power control} into a $ \rm{max} $ problem. A new form of the optimization is given in the following.	
	
	\begin{problem}\label{Intra-Block Power Control optimization}
		\begin{align}
			\mathop{\mathrm{maximize}}& \quad \beta_{L,g}[m] \nonumber \\
			s.t.&\quad R_{1,g}[m]=R_{2,g}[m]=\cdots=R_{L,g}[m] \nonumber\\
			&\quad \beta_{l,g}[m]\ge0\quad \forall l\in[1:L]
		\end{align} 
	\end{problem}
	
	By setting $ L=2 $, the relationship between $ \beta_{1,g}[m] $ and $ \beta_{2,g}[m] $ can be calculated through quadratic formula by solving $ R_{1,g}[m]=R_{2,g}[m] $. Then, the relationship between $ \beta_{1,g}[m] $ and $ \beta_{2,g}[m] $ in Eq.~\eqref{the relationship between factors} can be further obtained.
	
	\section{Proof of Corollary \ref{Limiting Rate of OMF-NOMA}}
	\label{Proof of Corollary Limiting Rate of OMF-NOMA}
	Let $ \frac{M}{K}=r $ and $\varpi_{l}=\mathsf{E}\left[\frac{|h_{\mathcal{I}_{Ml}}|^2}{|h|^2} \right]$.
	The limiting rate of NB-CoMAC in Theorem \ref{Older} is given as
	\begin{equation}
		\begin{split}
			R=&\dfrac{1}{B}\mathsf{E}\left[ \mathsf{C}^+\left( \dfrac{1}{M}+\dfrac{|h_{\mathcal{I}_{M}}|^2KP}{M\mathsf{E}\left[\dfrac{|h_{\mathcal{I}_{M}}|^2}{|h|^2} \right]}\right)\right]\\
			=&r\mathsf{E}\left[ \mathsf{C}^+\left( \dfrac{1}{rK}+\dfrac{\mathcal{N}_rP}{r\varpi_1}\right)\right]\\
			\stackrel{(a)}{=}&r\mathsf{C}^+\left( \dfrac{1}{rK}+\dfrac{\xi_{1-r}P}{r\varpi_1}\right),
		\end{split}
	\end{equation}
	where $ \mathcal{N}_r $  asymptotically follows the Gaussian distribution with mean $ \xi_{1-r} $ and variance $ \frac{r(1-r)}{Kf_{|h|^2}^2(\xi_{1-r})} $, $ F_{|h|^2}(\xi_{1-r})= 1-r$ \cite{wilks1948order} and condition $ (a) $ holds while the variance of $ \mathcal{N}_r  $ approaches 0.
	
	With similar steps, the limiting rate of WB-CoMAC in Theorem \ref{ESubPowerINC} is
	\begin{equation}
		\begin{split}
			R&=\dfrac{M}{KN}\mathsf{E}\left[\sum_{g=1}^{N} \mathsf{C}^+\left( \dfrac{N}{M}+ \dfrac{KP|h_{\mathcal{I}_{M}^g,g}|^2}{M\mathsf{E}\left[\dfrac{|h_{\mathcal{I}_{M}^g,g}|^2}{|h_{g}|^2} \right]}\right)\right]\\
			&=r\mathsf{C}^+\left( \dfrac{N}{rK}+\dfrac{\xi_{1-r}P}{r\varpi_1}\right).
		\end{split}
	\end{equation}
	
	And the limiting rate of NOMA-CoMAC with average power control in Corollary \ref{intra-block power alloaction} can be express as 
	\begin{equation}
		\resizebox{!}{!}{
			$
			\begin{split}
			R&=\dfrac{2}{BN}\mathsf{E}\left[\sum_{g=1}^{N}\mathsf{C}^{+}\left( \dfrac{N}{M}+\dfrac{2P\frac{K}{M}|h_{\mathcal{I}_{M}^g,g}|^2 }{\Gamma+\sqrt {
					\Gamma^{2}+4
					\,{ P\frac{K}{M}}\,{ |h_{\mathcal{I}_{M}^g,g}|^2}{ \varpi_{1,g}} }} \right)\right]\\
			&=2r\mathsf{E}\left[\mathsf{C}^{+}\left( \dfrac{N}{rK}+\dfrac{2P\mathcal{N}_r\mathcal{N}_{2r}}{r\Delta+\sqrt{(r\Delta)^2+4r\varpi_1P\mathcal{N}_r\mathcal{N}_{2r}^2}}\right)\right]\\
			&=2r\mathsf{C}^{+}\left( \dfrac{N}{rK}+\dfrac{2P\xi_{1-r}\xi_{1-2r}}{r\Delta_{\rm M} +\sqrt{(r\Delta_{\rm M})^2+4r\varpi_1P\xi_{1-r}\xi_{1-2r}^2}}\right),
			\end{split}
			$
		}
	\end{equation}
	where $ \Gamma=\frac{{ |h_{\mathcal{I}_{M}^g,g}|^2}}{{ |h_{\mathcal{I}_{2M}^g,g}|^2}}{ \varpi_{2,g}}+{ \varpi_{1,g} }$, $ \Delta=\varpi_1\mathcal{N}_{2r}+\varpi_2\mathcal{N}_{r} $ and $\Delta_{\rm M}=\mathsf{E}\left[\Delta \right]=\varpi_1\xi_{1-2r}+\varpi_2\xi_{1-r} $
	.

	\section{Proof of Corollary \ref{NOMA gain}}
	\label{Proof of Corollary NOMA gain}
	Recalling Problem \ref{optimization for intra-block power control} and setting $ L=2 $, the instantaneous computation rate is given as 
	\begin{equation}\label{key}
		R[m]=\dfrac{2M}{KN}\sum_{g=1}^{N}\min\left[ R_{1,g}[m],R_{2,g}[m]\right]
	\end{equation}
	with fixed power allocation factors $ \beta_{1,g}[m] $ and $ \beta_{2,g}[m] $.
	
	Since the analysis of outage probability can be seen as a single-input and multiple-output system approximately referring to \cite{tse2005fundamentals} in the OFDM-based system, we only aim at the outage probability of the rate in each sub-carrier $ P_{out}^{sub} $, and the outage probability $ P_{out} $ can be evaluated as
	\begin{equation}\label{overall outage probability}
	\resizebox{\linewidth}{!}{$
		\begin{aligned}
		P_{out}=&(P_{out}^{sub})^{N}\\
		=&(1-\Pr\left\lbrace R_{1,g}[m]>R^T,R_{2,g}[m]>R_{1,g}[m] \right\rbrace-\Pr\left\lbrace R_{2,g}[m]>R^T,R_{1,g}[m]>R_{2,g}[m] \right\rbrace)^N,
		\end{aligned}
		$}
\end{equation}
	where $ R^T $ is the target rate.
	
	In the following derivation, we will ignore $ [m] $ which stands for the $ m $-th OFDM symbol, since it has no influence on the result and makes derivation obscure. Hence, $ \bar{P}_{out}^{sub}=1-P_{out}^{sub} $ can be further expressed as
	\begin{equation}\label{NOMA Outage}
	\begin{split}
		\bar{P}_{out}^{sub}=\underbrace{\Pr\left\lbrace{\phi_1}<|h_{\mathcal{I}_{M}^g,g}|^2<\phi_2, |h_{\mathcal{I}_{2M}^g,g}|^2>\dfrac{\xi}{P}\right\rbrace}_{\alpha_1}+\underbrace{\Pr\left\lbrace|h_{\mathcal{I}_{M}^g,g}|^2>\phi_3,|h_{\mathcal{I}_{2M}^g,g}|^2>\dfrac{\varepsilon_{R^T}}{NP\beta_{2,g}}\right\rbrace}_{\alpha_2}
	\end{split}
\end{equation}
	where $ \phi_1= \frac{\varepsilon_{R^T}}{N\beta_{1,g}P}+\frac{\varepsilon_{R^T}\beta_{2,g}}{\beta_{1,g}}|h_{\mathcal{I}_{2M}^g,g}|^2 $, $\phi_2= \frac{\beta_{2,g}}{\beta_{1,g}}|h_{\mathcal{I}_{2M}^g,g}|^2+NP\frac{\beta_{2,g}}{\beta_{1,g}}|h_{\mathcal{I}_{2M}^g,g}|^4 $, $\phi_3= \frac{\beta_{2,g}}{\beta_{1,g}}|h_{\mathcal{I}_{2M}^g,g}|^2+NP\frac{\beta_{2,g}}{\beta_{1,g}}|h_{\mathcal{I}_{2M}^g,g}|^4 $, $ \xi=\frac{\varepsilon_{R^T}-1+\sqrt{\frac{4\varepsilon_{R^T}}{\beta_{2,g}}+(\varepsilon_{R^T}-1)^2}}{2N} $ and $ \varepsilon_{R^T}=2^{\frac{R^TKN}{2M}}-\frac{N}{M} $.
	
	As shown in \cite{yang2011order} and binomial theorem, the joint probability density function of the order statistics $ |h_{\mathcal{I}_{M}^g,g}|^2 $ and $| h_{\mathcal{I}_{2M}^g,g} |^2$ is 
	\begin{equation}\label{key}
	\resizebox{\linewidth}{!}{$
		\begin{split}
		f_{|h_{\mathcal{I}_{M}^g,g}|^2,| h_{\mathcal{I}_{2M}^g,g} |^2}(x,y)=\Theta\sum_{u_1=0}^{M-1}\sum_{u_2=0}^{K-2M}{{2M-1}\choose{u_1}}{{K-2M}\choose{u_2}}(-1)^{K-2M+u_1-u_2}\mathrm{Ex}_1(x)\mathrm{Ex}_2(y),
		\end{split}
		$}
\end{equation}
	where $ \Theta= \frac{K!}{(M-1)!(M-1)!(L-2M)!}$, $ \mathrm{Ex}_1(x)=e^{-(M+u_1)x} $ and $ \mathrm{Ex}_2(y)=e^{-(K-M-u_1-u_2)y} $.
	
	Since $ \bar{P}_{out}^{sub} $ can be divided into two parts, the first part $ \alpha_1 $ is evaluated as follows:
	\begin{equation}\label{key}
		\begin{split}
			\alpha_1=\Theta\sum_{u_1=0}^{M-1}\sum_{u_2=0}^{K-2M}{{M-1}\choose{u_1}}{{K-2M}\choose{u_2}}(-1)^{K-2M+u_1-u_2}\int_{\frac{\xi}{P}}^{\infty}\int_{\phi_1}^{\phi_2}\mathrm{Ex}_1(x)\mathrm{d}x\mathrm{Ex}_2(y)\mathrm{d}y.
		\end{split}
	\end{equation}
	And the second part can be evaluated as follows:
	\begin{equation}\label{key}
		\begin{split}
			\alpha_2=&\Theta\sum_{u_1=0}^{M-1}\sum_{u_2=0}^{K-2M}{{M-1}\choose{u_1}}{{K-2M}\choose{u_2}}(-1)^{K-2M+u_1-u_2}\int_{\frac{\varepsilon_{R^T}}{NP\beta_{2M}}}^{\infty}\int_{\phi_3}^{\infty}\mathrm{Ex}_1(x)\mathrm{d}x\mathrm{Ex}_2(y)\mathrm{d}y.
		\end{split}
	\end{equation}
	
	Substituting the above equations into Eq.~\eqref{overall outage probability}, the outage probability $ P_{out} $ in Eq.~\eqref{outage probability} can be obtained.
	
	In order to find the diversity gain, we focus on the case with high SNR, i.e., $ P\to\infty $. First of all, a more excat expression of $ \alpha_1 $ without the integral can be expressed as
	\begin{equation}\label{key}
		\resizebox{!}{!}{$
			\begin{split}
			\alpha_1=&\Theta\sum_{u_1=0}^{M-1}\sum_{u_2=0}^{K-2M}{{M-1}\choose{u_1}}{{K-2M}\choose{u_2}}(-1)^{K-2M+u_1-u_2}\\
			&\times\dfrac {1}{2\sqrt { ( M+{ u_1} ) D} ( M+{ u_1}
				)  (  ( -B+1 ) { u_1}+ ( -B+1 ) M-
				K+{ u_2} ) } \\
			&\times\left(  \left( {\rm erf} \left({\dfrac {
					( M+{ u_1} ) C+2\,DE ( M+{ u_1} ) +K-M-{
						u_1}-{ u_2}}{2\sqrt { ( M+{ u_1} ) D}}}\right)-1
			\vphantom{{{\rm e}^{{\dfrac { (  ( C-1 ) 
								{ u_1}+ ( C-1 ) M+K-{ u_2} ) ^{2}}{ 4( M+{
									u_1} ) D}}}}}
			\right)\right.\\
			&\times(  ( -B+1 ) { u_1}+ ( -B+1 ) M-
			K+{ u_2} ){{\rm e}^{{\dfrac { (  ( C-1 ) 
							{ u_1}+ ( C-1 ) M+K-{ u_2} ) ^{2}}{ 4( M+{
								u_1} ) D}}}}\sqrt {\pi}\\
			&\times\left.-2\,{{ e}^{ (  ( -B+1
					) { u_1}+ ( -B+1 ) M-K+{ u_2} ) E- ( 
					M+{ u_1} ) A}}\sqrt { ( M+{ u_1} ) D} \vphantom{{{ e}^{{\dfrac { (  ( C-1 ) 
								{ u_1}+ ( C-1 ) M+K-{ u_2} ) ^{2}}{ 4( M+{
									u_1} ) D}}}}}\right)	
			\end{split},
			$}
	\end{equation}
	where $ A=\frac{\varepsilon_{R^T}}{N\beta_{1,g}P} $, $ B=\frac{\varepsilon_{R^T}\beta_{2,g}}{\beta_{1,g}} $, $ C=\frac{\beta_{2,g}}{\beta_{1,g}} $, $ D=NP\frac{\beta_{2,g}}{\beta_{1,g}} $ and $ E=\frac{\xi}{P} $.
	With the case as $ P\to\infty  $, $ A\to 0 $, $ E\to 0 $ and $ D\to\infty $. Applying Taylor expansion of the exponential functions, $ \alpha_1 $ can be further given as
	\begin{equation}\label{alpha 1 form 2}
		\resizebox{!}{!}{$
			\begin{split}
			\alpha_1=&\Theta\sum_{u_1=0}^{M-1}\sum_{u_2=0}^{K-2M}{{M-1}\choose{u_1}}{{K-2M}\choose{u_2}}(-1)^{K-2M+u_1-u_2+1}\\
			&\times\dfrac {\left({\frac {
						\sqrt {\pi} \left(  \left( B-1 \right) { u_1}+ \left( B-1 \right) M-
						{ u_2}+K \right) }{\sqrt { \left( M+{ u_1} \right) D}}{\displaystyle\sum_{n=0}^{\infty}{{\dfrac { \left(  \left( C-1 \right) { u_1}+ \left( C-1 \right) M+
									K-{ u_2} \right) ^{2n}}{n!\left( 4 \left( M+{ u_1} \right) D\right) ^n}}}}}\right)}{ 2\left(  \left( B-1 \right) { u_1}+ \left( B-1 \right) M
				-{ u_2}+K \right)  \left( M+{ u_1} \right) }		
			\end{split}.
			$}
	\end{equation}
	By using the property of CDF of $| h_{\mathcal{I}_{2M}^g,g} |^2$ as $ F_{|h_{\mathcal{I}_{M}^g,g}|^2,| h_{\mathcal{I}_{2M}^g,g} |^2}(\infty,\infty) =1$ and the binomial theorem, \eqref{alpha 1 form 2} can be expressed as follows:
	
	\begin{equation}\label{key}
		\resizebox{!}{!}{$
			\begin{split}
			{\alpha _1} =& \frac{1}{2}\left( 1 + \Theta \sum\limits_{{u_1} = 0}^{M - 1} {	{M - 1}\choose{u_1}} {{( - 1)}^{K - 2M + {u_1} + 1}}\frac{{\sqrt \pi  \left( {\left( {B - 1} \right){u_1} + \left( {B - 1} \right)M - {u_2} + K} \right)}}{{\left( {\left( {B - 1} \right){u_1} + \left( {B - 1} \right)M - {u_2} + K} \right)\left( {M + {u_1}} \right)}}\right.\\
			&\times\left. \sum\limits_{n = 0}^\infty  \frac{1}{{n!{{\left( {4\left( {M + {u_1}} \right)D} \right)}^n}}}\sum\limits_{{u_2} = 0}^{K - 2M} {{{\left( { - 1} \right)}^{{u_2}}}}	{{K-2M}\choose{u_2}} \right. \\
			&\times\left.\sum\limits_{k = 0}^{2n - 1} {{2n - 1}\choose{k}} {{\left( {\left( {C - 1} \right){u_1} + \left( {C - 1} \right)M + K-u_2} \right)}^{2n - 1 - k}}{{\left( { - 1} \right)}^k}{{{{u_2}} }^k} \right)
			\end{split}.
			$}
	\end{equation}
	Recalling two sums of the binomial coefficients from \cite[Eq.~(0.153
	)]{gradshteyn2000table} and \cite[Eq.~(0.154
	)]{gradshteyn2000table}, we can have the following expression
	\begin{equation}\label{alpha 1 form 4}
		\resizebox{!}{!}{$
			\alpha_1\approx\dfrac{1}{2}-\Theta\sum_{u_1=0}^{M-1}{{M-1}\choose{u_1}}{{\dfrac { \left(  \left( C-1 \right) { u_1}+ \left( C-1 \right) M+
						K-{ u_2} \right) ^{2(K-2M)}}{(K-2M)!\left( 4 \left( M+{ u_1} \right) D\right) ^{(K-2M)}}}},
			$}
	\end{equation}
	since the factor with $ u_2^k $, $ k<(K-2M) $ is equal to zero by using the above equations. Furthermore, the terms with $ u_2^k $, $ k>(K-2M) $ can be also removed as the dominant factor is $ k=K-2M $. Following steps similar to the ones for obtaining Eq.~\eqref{alpha 1 form 4}, we can prove that the dominant factor is also $ k=K-2M $ in $ \alpha_2 $. In conclusion, the diversity gain is obtained.
	\bibliography{NOAH}
	\bibliographystyle{IEEEtran}
\end{document}